\documentclass[AMA,STIX1COL]{WileyNJD-v2}
\usepackage{amsmath}
\usepackage{bm, icomma}
\usepackage{setspace}

\articletype{Research Article}

\begin{document}

\title{Bayesian unanchored additive models for component network meta-analysis}

\author[1]{Augustine Wigle*}

\author[1]{Audrey B{\'e}liveau}

\authormark{WIGLE and B{\'E}LIVEAU}

\address[1]{\orgdiv{Department of Statistics and Actuarial Science}, \orgname{University of Waterloo}, \orgaddress{\state{Ontario}, \country{Canada}}}

\corres{*Augustine Wigle, Department of Statistics and Actuarial Science, University of Waterloo, Ontario, Canada. \email{amhwigle@uwaterloo.ca}}

\abstract[Summary]{Component Network Meta-Analysis (CNMA) models are an extension of standard Network Meta-Analysis (NMA) models which account for the use of multicomponent treatments in the network. This paper contributes innovatively to several statistical aspects of CNMA. First, by introducing a unified notation, we establish that currently available methods differ in the way they assume additivity, an important distinction that has been overlooked so far in the literature. In particular, one model uses a more restrictive form of additivity than the other which we term an anchored and unanchored model, respectively. We show that an anchored model can provide a poor fit to the data if it is misspecified. Second, given that Bayesian models are often preferred by practitioners, we develop two novel unanchored Bayesian CNMA models presented under the unified notation. An extensive simulation study examining bias, coverage probabilities and treatment rankings confirms the favourable performance of the novel models. This is the first simulation study to compare the statistical properties of CNMA models in the literature. Finally, the use of our novel models is demonstrated on a real dataset, and the results of CNMA models on the dataset are compared.}

\doublespacing


\thispagestyle{empty}
\noindent \large This is the peer reviewed version of the following article: 

Wigle A, B{\'e}liveau A. Bayesian unanchored additive models for component network meta-analysis. \textit{Statistics in Medicine} 2022;41(22): 4444-4466,

\noindent which has been published in final form at \hyperlink{10.1002/sim.9520}{10.1002/sim.9520}. This article may be used for non-commercial purposes in accordance with Wiley Terms and Conditions for Use of Self-Archived Versions. This article may not be enhanced, enriched or otherwise transformed into a derivative work, without express permission from Wiley or by statutory rights under applicable legislation. Copyright notices must not be removed, obscured or modified. The article must be linked to Wiley’s version of record on Wiley Online Library and any embedding, framing or otherwise making available the article or pages thereof by third parties from platforms, services and websites other than Wiley Online Library must be prohibited.

\setcounter{page}{0}

\clearpage

\maketitle

\newcommand{\logit}{\text{logit}}
\newcommand{\logOR}{\text{log OR}}
\newcommand{\expit}{\text{logit}^{-1}}
\newcommand{\MVN}{\text{MVN}}
\newcommand{\Unif}{\text{Unif}}
\newcommand{\Bin}{\text{Binomial}}
\newcommand{\N}{\text{N}}
\newcommand{\dbar}[1]{d^*_{#1}}
\newcommand{\deltaij}{\delta_{i,ref, t_{ij}}}
\newcommand{\deltaii}{\delta_{i,ref, t_{i1}}}
\newcommand{\deltaik}{\delta_{i,ref,t_{ik}}}
\newcommand{\deltaiai}{\delta_{i,ref,t_{ia_i}}}
\newcommand{\deltaref}[1]{\delta_{i,ref,#1}}
\newcommand{\deltabar}[1]{\delta^*_{i,#1}}
\newcommand{\tij}{t_{ij}}
\newcommand{\tii}{t_{i1}}
\newcommand{\probi}{\bm{p}_i}
\newcommand{\pp}{\bm{p}}
\newcommand{\aali}{\bm{\alpha}_i}
\newcommand{\dda}{\dd^{(1)}}
\newcommand{\Aistar}{\bm{V}_i^{(1)}}
\newcommand{\epsi}{\bm{\epsilon}_i}
\newcommand{\epsistar}{\bm{\epsilon}_i^*}
\newcommand{\Ai}{\bm{V}_i}
\newcommand{\Bi}{\bm{U}_i}
\newcommand{\one}{\bm{1}_{1\times a_i}}
\newcommand{\dd}{\bm{d}}
\newcommand{\Omg}{\bm{\Omega}}
\newcommand{\dref}[1]{d_{ref,#1}}
\newcommand{\Deltaij}{\Delta_{i,j}}
\newcommand{\yy}{\bm{y}}
\newcommand{\yyi}{\bm{y}_i}
\newcommand{\XX}{\bm{X}}
\newcommand{\AAA}{\bm{V}}
\newcommand{\ZZ}{\bm{Z}}
\newcommand{\VV}{\bm{V}}
\newcommand{\tth}{\bm{\theta}}
\newcommand{\bbet}{\bm{\beta}}
\newcommand{\aal}{\bm{\alpha}}
\newcommand{\mmu}{\bm{\mu}}
\newcommand{\dvec}{\dd}
\newcommand{\dbarvec}{\bm{d^*}}
\newcommand{\BB}{\bm{U}}
\newcommand{\CC}{\bm{C}}
\newcommand{\WW}{\bm{W}}
\newcommand{\E}{\textbf{E}}
\newcommand{\sumtij}{\sum_{k \in \tij}}
\newcommand{\dabc}[2]{d^{(#1)}_{#2}}
\newcommand{\hatdabc}[2]{\hat{d}^{(#1)}_{#2}}
\newcommand{\epss}{\bm{\epsilon}_1}
\newcommand{\epsss}{\bm{\epsilon}}
\newcommand{\zero}{\bm{0}}
\newcommand{\bigS}{\bm{S}}
\newcommand{\nij}{n_{ij}}

\newcommand{\drela}[1]{d_{A,#1}}
\newcommand{\drelb}[1]{d_{B,#1}}
\newcommand{\drelc}[1]{d_{C,#1}}
\newcommand{\drelx}[1]{d_{X,#1}}
\newcommand{\dxwrt}[2]{d_{X,#1}^{(#2)}}
\newcommand{\dywrt}[2]{d_{Y,#1}^{(#2)}}
\newcommand{\dwrt}[2]{d_{#1}^{(#2)}}

\section{Introduction}

Network Meta-Analysis (NMA) is a statistical methodology used to estimate the relative effect of multiple competing treatments based on data from Randomized Controlled Trials (RCTs). The use of complex interventions in NMAs has been rising, and methodological advances related to complex interventions are increasing. \citep{petropoulou2021review} Complex interventions refer to treatment protocols which are composed of multiple components, and are also referred to as multicomponent treatments. For example, one treatment for depression may be composed of anti-depressant drug A together with counselling. There may be arms within the network which employ these components alone, or in a different combination, such as anti-depressant drug A alone or anti-depressant drug B with counselling. For estimating the effect of components in complex interventions, standard NMA models have been extended \citep{petropoulou2021review, welton2009mixed, rucker2019network, efthimiou2022bayesian}; such models are referred to as Component NMA (CNMA) models.

In a recent review of existing meta-analytic methods for complex interventions, Petropoulou et al. 2021 \citep{petropoulou2021review} recommended the use of additive CNMA models (when the assumptions are satisfied) over other types of CNMA models. Additive CNMA models were preferred because they allow for comparison of relative efficacy of all possible combinations of components, and because of the availability of code and R packages which facilitate their implementation. Existing additive CNMA models have been applied in practice by e.g. Mills et al. (2011, 2012) \citep{mills2011pharmacotherapies, mills2012calculating}; Madan et al. (2014) \citep{madan2014synthesis}; Caldwell and Welton (2016) \citep{caldwell2016approaches}; Pompoli at al. (2018) \citep{pompoli2018dismantling}; and Freeman et al. (2018).\citep{freeman2018component}

There are two established additive CNMA methods; one proposed by Welton et al. (2009) \citep{welton2009mixed} and another proposed by R{\"u}cker et al. (2020).\citep{rucker2019network} The model of Welton et al. (2009) \citep{welton2009mixed} uses Bayesian inference and takes trial data summarised at the arm level. The approach of R{\"u}cker et al. (2020) \citep{rucker2019network} is a frequentist model which uses data summarised at the contrast level. The two methods also differ in the way multi-arm trials are accounted for. Recently, several Bayesian CNMA models were proposed by Efthimiou et al. (2022).\citep{efthimiou2022bayesian} Since our paper was written before publication of Efthimiou et al. (2022)\citep{efthimiou2022bayesian}, the terms ``existing" or ``established" methods in this paper refer to the methods of Welton et al. (2009) \citep{welton2009mixed} and R{\"u}cker et al. (2020).\citep{rucker2019network} Efthimiou et al. (2022)\citep{efthimiou2022bayesian} focuses on extending CNMA models to allow for the use of individual patient data and facilitating selection of component interactions. They describe their model III as Welton et al.'s (2009) \citep{welton2009mixed} model, however we note (from inspecting their model formulation and code) that it is not exactly Welton et al.'s (2009) \citep{welton2009mixed} model, but rather analogous to a model we propose in this paper in Subsection \ref{sec:condat}. Our paper clarifies the distinction between these models.

Both Frequentist and Bayesian approaches to NMA are commonly used.\citep{dias2014generalised, dias2018network, rucker2012network} However, Bayesian methods are often preferred by NMA practitioners in decision-making contexts because they present several advantages.\citep{dias2018network} First, Bayesian approaches are flexible and can be adapted to a variety of data types. Bayesian approaches that use arm-level data can even accommodate binomial data where some trials have zero counts without requiring adjustment. Second, Bayesian approaches automatically account for the uncertainty in the estimation of heterogeneity in random-effects models, whereas a frequentist approach estimates the heterogeneity parameter and then treats it as fixed and known. The uncertainty in the heterogeneity from a Bayesian approach is propagated to the posterior distributions of the relative effects, allowing for a better reflection of the true uncertainty in the estimates. Lastly, Bayesian approaches naturally lend themselves to health-economic decision modeling (see p. 177, Dias et al. 2018 \citep{dias2018network}).

There has been limited work done assessing the existing CNMA models. Thorlund and Mills (2012) \citep{thorlund2012stability} investigated the performance of the standard Bayesian NMA model compared to the additive CNMA model of Welton et al. (2009) \citep{welton2009mixed} in a simulation study. R{\"u}cker et al. (2020) \citep{rucker2020component} compared the frequentist CNMA model to a matching method on a dataset of multiple myeloma. However, no direct comparison of these CNMA methods has been made so far in the literature. The two models use different notation for different elements of the models, which makes comparison difficult. 

In the two established CNMA methods, the simplest model available is an additive model, where it is assumed that the effects of components are additive when they are combined in multicomponent treatment protocols. 
In Welton et al. (2009) \citep{welton2009mixed}, this is described mathematically as 
\begin{equation}
    d_{A,X} = \sum_{C \in X} d_{A,C}\label{eq:additive}
\end{equation}
for all multicomponent treatments $X$ in the network, where $C$ indexes the component treatments of $X$ and $d_{A,A'}$ represents the relative effect of treatment $A'$ compared to treatment $A$. We refer to \eqref{eq:additive} as the additivity equation and we term the treatment $A$ in equation \eqref{eq:additive} the anchor treatment. Simply put, equation \eqref{eq:additive} indicates that additivity is implemented on the relative effect scale, and specifically on the scale where effects are relative to the anchor treatment A. We introduce the new term ``anchor treatment" because the role of this treatment is distinct from the reference treatment or baseline treatment in an analysis, which is emphasized in Subsection \ref{sec:importance}.
While the inclusion of additional interaction terms in equation \eqref{eq:additive} can allow for a relaxation of additivity, in this work, we focus on the additivity as presented in \eqref{eq:additive} to simplify our presentation.

This paper contributes to several aspects of additive CNMA models. First, by introducing a new unified notation for the models, we are able to compare the existing CNMA methods of Welton et al. (2009) \citep{welton2009mixed} and R{\"u}cker et al. (2020) \citep{rucker2019network} qualitatively. Our work identifies a key difference between these two approaches which has been overlooked until now: the approaches differ in the way the anchor treatment is chosen. In particular, the additive CNMA model of Welton et al. (2009) \citep{welton2009mixed} relies on the modeller specifying the anchor treatment which is used in calculating additive effects, whereas the model of R{\"u}cker et al. (2020) \citep{rucker2019network} does not require the modeller to choose a specific anchor. Rather, it flexibly selects the anchor in a data-driven process. We call the first approach an anchored model and the second approach an unanchored model. We argue that the first approach is a more restrictive one which can be problematic if the modeller-imposed anchor treatment is misspecified.

Second, given the desirable flexibility of unanchored models and the popularity of Bayesian methods, we develop two unanchored Bayesian additive CNMA models as an alternative to the established approaches. While the frequentist approach of R{\"u}cker et al. (2020)\citep{rucker2019network} can only analyse contrast-level data, we propose two models, one to analyse contrast-level data and one to analyse arm-level data. Further, to facilitate implementation of our models by practitioners, we demonstrate the models on a real dataset and provide code for their computational implementation. 

The third contribution is an empirical comparison of the performance of our novel models with the existing models using an extensive simulation study. This is the first simulation study to compare the existing CNMA models. The study shows that serious biases can occur when using an anchored model. The study also demonstrates the robustness of all the unanchored models, including our two novel Bayesian models.  

The paper is structured as follows: In Section \ref{sec:background}, the additive CNMA models of Welton et al. (2009) \citep{welton2009mixed} and R{\"u}cker et al. (2020) \citep{rucker2019network} are compared using unified notation, and each model is identified as either anchored or unanchored. Section \ref{sec:anchor} explains the role of the anchor treatment in CNMA models and its impact on model fit, and defines anchored and unanchored models. The type of additivity used in existing CNMA models is also identified in this section. In Section \ref{sec:newmodels}, two Bayesian unanchored additive CNMA models are introduced. Section \ref{sec:sims} contains a simulation study comparing the four additive CNMA models discussed in this paper. In Section \ref{sec:casestudy}, the models are applied to a real dataset. The paper concludes with a discussion and future directions for this work in Section \ref{sec:disc}. 

\section{Comparison of Existing Models}\label{sec:background}

In this section, we present the existing models of Welton et al. (2009) \citep{welton2009mixed} and R{\"u}cker et al. (2020).\citep{rucker2019network} We unify their presentation using a new matrix notation and design matrices, which are ultimately helpful in comparing them in Subsection \ref{sec:anchored}. The notation is summarised in Table \ref{tab:summary}.

\subsection{Bayesian CNMA Model of Welton et al. (2009) \citep{welton2009mixed}} \label{sec:welton}

Welton et al. (2009) \citep{welton2009mixed} proposed a Bayesian method for analysing multicomponent treatments and the effects of their components using data summarised at the arm level. This model is typically fit using Markov Chain Monte Carlo (MCMC) sampling via BUGS software. BUGS code corresponding to the model as originally presented is available in Welton et al. (2009).\citep{welton2009mixed} Below we introduce the fixed and random effects models of Welton et al. (2009) \citep{welton2009mixed} using a matrix notation, which facilitates comparison to other models.

In this model, a reference treatment must be chosen. The reference treatment is the treatment labelled as ``1''. If there are $c$ components in the network, the model estimates $c-1$ relative component effects in which the chosen reference treatment is the comparator. Elements of the model whose structure depends on the treatment chosen to be treatment 1 are written with a superscript $(1)$.

Suppose we have $I$ RCTs, indexed with $i$, each with a total of $a_i$ arms. Assuming the outcome of interest is binary, such as death, then the trial data are aggregated as the counts of the outcomes in arm $j$ of study $i$, $r_{ij}$, and we assume that
\begin{equation}\label{eq:bin}
r_{ij} \mid p_{ij} \sim \Bin(n_{ij}, p_{ij})
\end{equation}
given $n_{ij}$, the total number of subjects observed in arm $j$ of study $i$, and $p_{ij}$, the probability of an outcome in the respective arm and study.


\begin{table}[]
    \centering
    \caption{Summary of the model elements used in the new unified notation for additive CNMA models in this paper.}
    \begin{tabular}{c|c|l}
        Model Element & Role & Description \\
        \hline
        $\Bi$, ${\Bi}^*$ & Design Matrix & Form contrasts between arms - only for contrast-based data \\
        $\Ai$, $\Aistar$ & Design Matrix & Form component structure of treatment arms \\
        \hline
        $\alpha_i$ & Parameter & Represents trial-specific response to the anchor treatment - only for\\
        & &  arm-level data \\
        $\sigma^2$ & Parameter & Represents between-trial heterogeneity in random effects models \\
        $\dda$ & Parameter & Vector of relative effects for components $2$, $\dots$, $c$ where treatment 1 \\
        & & is the comparator - only for anchored model \\
        $\dd$ & Parameter & Vector of relative effects for components $1$, $\dots$, $c$ where the unknown \\
        & & anchor is the comparator - only for unanchored models \\
        \hline
        $\one$ & Structural Element & Vector of ones of length $a_i$ - only for arm-level data\\
        $\Sigma_i$, $\Sigma_i^*$ & Structural Element & Square matrix of ones and $1/2$'s which describes correlation \\
        & & structure - only for Bayesian random effect models \\
        \hline
        $\yyi$, $\yyi^*$ & Data & Vector of contrast data - only for contrast-based data \\
        $r_{ij}$, $\nij$ & Data & Binomial counts and sample sizes for $j^{\text{th}}$ arm in trial $i$ - only \\
        & & for arm-level data \\
        $\Omg_i$ & Covariance Matrix & Covariance matrix incorporating uncertainty in data and from \\
        & & between-trial heterogeneity - only for frequentist model \\
        $\bigS_i^*$ & Covariance Matrix & Covariance matrix quantifying uncertainty in data - only for \\
        & & Bayesian contrast-based model \\
        \hline
        $*$ & Notation & Indicates elements of the model which are of reduced dimension due\\
        & &  to the inclusion of $a_i-1$ independent contrasts from multi-arm \\
        & & trials instead of $\binom{a_i}{2}$ dependent contrasts \\
        $(1)$ & Notation & Indicates elements of the model which are of reduced dimension \\
        & & because the model is anchored, and therefore the elements structure\\
        & &  is also dependent on which treatment was chosen as the anchor \\
        & & (treatment 1)
        
    \end{tabular}
    \label{tab:summary}
\end{table}

Let $\probi = (p_{i1}, \dots, p_{ia_i})'$ be the vector of binomial probabilities in trial $i$.  Without loss of generality, assume that if trial $i$ employed the reference treatment it is in the arm labelled 1. Then $\logit(\probi)$, the vector of the log-odds of outcomes in the arms of study $i$, can be modelled using
\begin{equation}
    \logit(\probi) = \alpha_i\one + \Aistar \dda + \epsi \label{eq:weltonmodel}
\end{equation}
where $\alpha_i$ is the baseline response in study $i$ to the reference treatment (the treatment labelled as ``1"); the vector $\one$ is a vector of ones of length $a_i$; $\dda = (d_{1,2}, \dots, d_{1,c})'$ is a vector of relative effects of length $c-1$ for the unique components, relative to the reference treatment; $\Aistar$ is a design matrix with dimension $a_i \times (c-1)$ where $V^{(1)}_{i,jk} = 1$ if arm $j$ involves component $k+1$ and $0$ otherwise. That is, if arm $j$ employs only the reference treatment, the row sum of $\Aistar$ is zero, otherwise the row sum will equal the number of components employed in the treatment protocol of arm $j$. Multiplying the $j^{\text{th}}$ row of $\Aistar$ by $\dda$ results in equation \eqref{eq:additive}, where $A$ is the treatment labelled as 1 and $B$ is the treatment protocol employed in arm $j$.

Further, $\epsi\sim \N(\zero, \sigma^2\Sigma_i)$, where 
\begin{equation}
    \Sigma_i = \begin{pmatrix}
    1 & 1/2 & \hdots & 1/2 \\
    1/2 & 1 & \hdots & 1/2 \\
    \vdots & \hdots & \ddots &\vdots \\
    1/2 & \hdots & \hdots & 1
    \end{pmatrix}_{(a_i) \times (a_i)}, \label{eq:epsidist}
\end{equation}
if treatment 1 is not used in trial $i$, otherwise $\Sigma_i$ has the same definition but with zeros in the first row and column (i.e. there is no heterogeneity in the effect of the reference treatment).  The parameter $\sigma$ represents the degree of between-study heterogeneity. The model as presented is called a random effects model. To obtain a fixed effect model, $\sigma$ is set to zero, which is equivalent to omitting $\epsi$ from equation \eqref{eq:weltonmodel}.

Independent diffuse normal priors are used for the baselines $\alpha_i$ and the relative effects $\dda$, such as $\N(\zero, 1000 \bm{I})$. A flat uniform prior is used for $\sigma$, where the upper limit is chosen based on the scale of the data.

Estimation is performed by using MCMC sampling. A posterior sample is obtained, usually by using BUGS.\citep{welton2009mixed} A point estimate for $d_{1,k}$, $k = 2, \dots, c$ and $\sigma$ can then be obtained by taking the mean or the median of the posterior samples. Multicomponent treatment effects can be obtained via the additivity equation \eqref{eq:additive}. The relative effect of any two treatments $k$ and $l$ can then be determined via the consistency equation,
\begin{equation}
    \hat d_{k,l} = \hat d_{1,l}-\hat d_{1,k} \label{eq:consistency}
\end{equation}
\citep{dias2014generalised}. Credible intervals can also be obtained using the appropriate quantiles of the posterior samples, for example, 2.5\% and 97.5\% for a 95\% credible interval.

\subsection{Frequentist CNMA Model of R{\"u}cker et al. (2020) \citep{rucker2019network}}\label{sec:rucker}

R{\"u}cker et al. (2020) \citep{rucker2019network} proposed a frequentist approach to CNMA which takes contrast-based summary data. The R package netmeta implements this method and is freely available.\citep{netmeta} We introduce their fixed and random effects models using unified notation which is summarised in Table \ref{tab:summary}.

A study with $a_i$ arms produces $\binom{a_i}{2} = m_i$ comparisons, where only $a_i-1$ of those comparisons are independent.\citep{rucker2014reduce}
 In the frequentist approach of R{\"u}cker et al. (2020),\citep{rucker2019network} when multi-arm studies are included, all $m_i$ observed contrasts are included, which differs from the other contrast-based Bayesian models where only the $a_i-1$ independent contrasts are included for multi-arm trials.\citep{welton2009mixed, dias2014generalised}

We introduce data and design matrices at the trial level to facilitate comparison to the model of Welton et al. (2009).\citep{welton2009mixed} Let the observed $m_i$ comparisons for study $i$ be denoted by $\yyi = (y_{i1}, \dots, y_{i m_i})'$ with associated standard errors $SE(y_{ij})$, $i = 1, \dots, I$ and $j = 1, \dots, m_i$. The pairwise comparisons could be in the form of log odds ratios, mean differences, or some other suitable measure.\citep{rucker2019network} The model is given by
\begin{equation}
\yyi \sim \N(\mmu_i, \Omg_i),
\end{equation}
with
\begin{equation}
\mmu_i = \Bi \Ai\dd
\end{equation}
where $\dd$ is a vector of length $c$ which contains ``treatment effects" for each component from $1$ to $c$ \citep{rucker2019network}; 
$\Bi$ is a $m_i \times a_i$ design matrix which creates a vector of $m_i$ contrasts from the $a_i$ arms in study $i$. In particular, $U_{i,jk} = 1$ if the $k^\text{th}$ arm is the minuend of the $j^\text{th}$ contrast, $-1$ if the $k^\text{th}$ arm is the subtrahend in the $j^\text{th}$ contrast, and $0$ otherwise.
$\Ai$ is a design matrix with dimension $a_i \times c$ where $V_{i,jk} = 1$ if arm $j$ involves component $k$ and zero otherwise. 

Note that in this model, we use the notations $\dd$ and $\Ai$ with no superscript to emphasize the fact that these quantities do not depend on a chosen anchor treatment, and thus are of full dimension, involving all $c$ unique treatment components. See Section \ref{sec:anchored} for an illustrative example.

$\Omg_i$ is a diagonal covariance matrix that reflects both the standard errors in the data, $SE(y_{ij})$, and, for a random-effects model, the degree of between-study heterogeneity estimated using a method-of-moments. Between-study heterogeneity can be omitted for a fixed effects model. We refer the reader to R{\"u}cker and Schwarzer (2014) \citep{rucker2014reduce} and R{\"u}cker et al. 2020 \citep{rucker2019network} for details on the specification of $\Omg_i$.

 To obtain a regression estimate of  $\dvec$, the model can be rewritten by letting $\yy = (\yy_1', \dots, \yy_I')'$, $\Omg$ be a block-diagonal matrix with $(\Omg_1, \dots, \Omg_I)$ along the diagonal, and
 \begin{equation}
     \XX = \begin{pmatrix}
     \BB_1 \AAA_1 \\
     \vdots \\
     \BB_I \AAA_I
     \end{pmatrix}.
 \end{equation}
We can then write
\begin{equation}
\yy \sim N(\XX \dvec, \Omg), \label{eq:ruckermodel}
\end{equation}
where $\yy$ and $\XX$ are the same as those defined in R{\"u}cker et al. (2020).\citep{rucker2019network}
Weighted least squares is used to estimate $\dd$, that is
\begin{equation}
    \widehat{\dvec} = (\XX^T \Omg^{-1} \XX)^+\XX^T \Omg^{-1} \yy, \label{eq:betaest}
\end{equation}
where $(\XX^T \Omg^{-1} \XX)^+$ is the Moore-Penrose generalized inverse of $\XX^T \Omg^{-1} \XX$.\citep{rucker2019network,rucker2014reduce}

To estimate any contrast, the appropriate contrast of relative effects in $\hat \dd$ can be taken, by applying additivity (equation \eqref{eq:additive}) and the consistency equations (equation \eqref{eq:consistency}). For example, if the relative effect of treatment $1+2$ compared to $3$ was desired, it would be calculated as
\begin{equation}
    \hat d_{3,1+2} = \hat d_{1} + \hat d_{2} - \hat d_{3},
\end{equation}
where $\hat d_{k}$ represents the $k^\text{th}$ entry in the vector $\hat \dd$.

\section{Anchored and Unanchored Models}\label{sec:anchor}

Recall that additivity is defined by equation \eqref{eq:additive}, and we term the treatment to which the effects are relative in this equation the anchor treatment. In Subsection \ref{sec:importance}, we show that the anchor treatment determines the relative effect parameters and different anchors correspond to different parameter values. We also show how the concept of the anchor treatment and the reference treatment differ. The purpose of this section is to explain the role of the anchor treatment and demonstrate that it is an important parameter of the model. In Subsection \ref{sec:anchored}, we introduce the terminology of anchored and unanchored models and identify which approach is used in the existing CNMA models reviewed in Section \ref{sec:background}. We also argue that anchored models are more prone to misspecification.

\subsection{Role of the Anchor Treatment}\label{sec:importance}

The anchor treatment defines the scale on which the additivity equation is applied.  Different forms of additivity are possible based on which treatment is the anchor. Furthermore, different anchors are not equivalent. For example, if A is the true anchor treatment, then the multicomponent relative effect parameters are defined according to $d_{A,X} = \sum_{C \in X} d_{A,C}$ for all multicomponent treatments $X$. However, this does not imply that $d_{B,X}$ is equal to $\sum_{C \in X} d_{B,C}$. In fact, if treatment B was the true anchor and thus multicomponent relative effects were determined according to $d_{B,X} = \sum_{C \in X} d_{B,C}$, this would represent a different set of true relative effects and thus a different data-generating mechanism. Here recall that although the additivity equation only allows for the calculation of effects relative to one treatment, consistency allows for the derivation of effects relative to any comparator. That is, given a set of effects relative to treatment $A$, $d_{A,k}$ for all treatments $k$, $d_{B,k}$ can be determined according to $d_{B,k} = d_{A,k} - d_{A,B}$.\citep{dias2014generalised} In Appendix \ref{sec:unique}, we give a rigorous proof showing that additivity can hold for only one anchor treatment at once. Therefore, the anchor treatment uniquely defines the form of additivity used to compute relative effects for multicomponent treatments, and different anchors are not equivalent. This is important to emphasize as it has been overlooked so far in the literature.


A further point worth emphasizing is the distinction between the anchor treatment and the reference treatment. It is difficult to differentiate between these two concepts, because in the existing Bayesian CNMA approach, we are restricted to using the same treatment as the anchor and the reference treatment (see following section). Despite this, they play distinct roles which is why we refer to them using different terms. While it is true that using any treatment as the reference treatment is equivalent due to the consistency equations, i.e.  
\begin{equation*}
d_{A,C+D} = d_{X,C+D} - d_{X,A}  = d_{Y,C+D} - d_{Y,A},  
\end{equation*}
it is not true that if
\begin{equation*}
d_{A,C+D} = d_{X,C} + d_{X,D} - d_{X,A}
\end{equation*}
(anchor is X and reference is X) , then 
\begin{equation*}
d_{A,C+D} = d_{Y,C} + d_{Y,D} - d_{Y,A}
\end{equation*}
(anchor is Y and reference is Y). Similarly, it is also not true that if 
\begin{equation*}
d_{A,C+D} = d_{X,C} + d_{X,D} - d_{X,A}
\end{equation*}
(anchor is X and reference is X) then
\begin{equation*}
    d_{A,C+D} = d_{Y,C} + d_{Y,D} - d_{X,A} 
\end{equation*}
(anchor is Y and reference is X). Thus if the true anchor was treatment X but the model used treatment Y as the anchor, the model would be misspecified. The choice of the reference treatment used to express relative effects is irrelevant because of the consistency equations. This is not true of the anchor treatment, as different anchors represents different data-generating mechanisms.

\subsection{Anchored and Unanchored Models}\label{sec:anchored}

Now that we have clarified the role of the anchor treatment, it is important to consider the implications for analysis. In additive CNMA models, an additivity equation is used to calculate multicomponent treatment effects, which means that the anchor treatment must be determined or estimated. However, the way the anchor is estimated in existing models has been overlooked. Due to a lack of discussion on this topic in the literature, it is unlikely that the true anchor treatment would be known. Further, if an anchor was specified in the analysis which is different from the true anchor, the model will be misspecified which could introduce bias to the estimated treatment effects and treatment rankings. For this reason, models which do not require the anchor to be specified a priori are attractive. Appendix \ref{sec:impact} includes more detailed examples of how the anchor treatment used in an analysis impacts the model and validity of results if it is misspecified.

In the models presented in Subsections \ref{sec:welton} and \ref{sec:rucker}, we argue that the anchor treatment is chosen in two different ways.
It is either specified by the modeller a priori or estimated by the model. The first method corresponds to assuming that a particular treatment in the network is the true anchor, whereas the second method corresponds to assuming that additivity holds for some unknown true anchor which is determined during model fitting. We define anchored models as models where the anchor is specified. Correspondingly, we define unanchored models as models where the anchor is estimated using the data.

Here, an analogy to simple linear models is useful. The true intercept of a linear model is analogous to the true anchor treatment in an additive CNMA model. Anchored models correspond to models which are fit assuming that the intercept is equal to zero. Unanchored models correspond to simply assuming that there is an intercept and estimating its value. Unless there is strong evidence from previous data or a physical reason why the true intercept is zero, we would prefer to estimate the intercept in fitting the model because if the intercept is misspecified it could bias the results of the model. Similarly, in the absence of problem-specific knowledge to justify choosing a particular anchor, it is likely safer to allow the anchor to be estimated using the data.

To illustrate whether existing CNMA approaches are anchored or unanchored, we give an illustrative example. Consider a network with components placebo, treatment A, and treatment B. Let us examine the design matrices and resulting parameters for a two-arm study with placebo in one arm and treatment A in the second arm. If we use the model of Welton et al. from Subsection \ref{sec:welton}, we must choose which treatment to use as treatment 1. If placebo is chosen as treatment 1, the design matrix $\Aistar$ is as follows:
\begin{equation}
\Aistar = \begin{pmatrix}
0 & 0 \\ 1 & 0
\end{pmatrix},
\end{equation}
and the resulting vector $\logit(\pp_i)$ is
\begin{equation}
\begin{pmatrix}
\alpha_i \\
\alpha_i
\end{pmatrix} + 
\begin{pmatrix}
0 & 0 \\ 1 & 0
\end{pmatrix}
\begin{pmatrix}
d_{1,A} \\
d_{1,B}
\end{pmatrix} = 
\begin{pmatrix}
\alpha_i + 0 \\
\alpha_i + d_{1,A}
\end{pmatrix}.
\end{equation}
This implies that $\alpha_i$ is the response to the placebo (treatment 1), and the elements of $\dda$ must be relative to placebo. In contrast, if treatment A is chosen to be treatment 1, we would obtain
\begin{equation}
\begin{pmatrix}
\alpha_i + d_{1,Placebo} \\
\alpha_i + 0
\end{pmatrix},
\end{equation}
showing that in general $\alpha_i$ is the response to whatever is chosen as treatment 1 and $\dda$ is a vector of effects relative to treatment 1. Whichever treatment is chosen to be treatment 1 is forced to be the anchor in the analysis because multicomponent treatment effects are computed as sums of the elements of $\dda$. The structure of the design matrix in this model assumes that a particular treatment is the anchor, thus the model of Welton et al. (2009) \citep{welton2009mixed} is an anchored model.

On the other hand, consider using the model from Subsection \ref{sec:rucker} proposed by R{\"u}cker et al. (2020).\citep{rucker2019network} The design matrices are as follows:
\begin{equation}
\Bi = \begin{pmatrix}
-1 & 1 
\end{pmatrix}; \hspace{0.2cm} \Ai = \begin{pmatrix}
1 & 0 & 0 \\
0 & 1 & 0
\end{pmatrix};
\end{equation}
The resulting parameter contrast is
\begin{equation}
\Bi\Ai\dd = \begin{pmatrix}
-1 & 1 
\end{pmatrix}\begin{pmatrix}
1 & 0 & 0 \\
0 & 1 & 0
\end{pmatrix} \begin{pmatrix}
d_{k,Placebo} \\
d_{k,A} \\
d_{k,B}
\end{pmatrix} = d_{k,A} - d_{k,Placebo},
\end{equation}
where $k$ represents the anchor. Multicomponent treatment effects are sums of the elements of $\dd$, which are relative to some unspecified treatment. The unspecified treatment is chosen based on what gives the best fit to the data. Therefore, the anchor treatment is estimated during model fitting and not specified by the modeller, unlike in the previous model. Thus, this model is an unanchored model.

A key difference between the anchored and unanchored models is in the dimension of the relative effect parameter vector. In the anchored model, the relative effect parameter vector $\dda$ is of length $c-1$. In the unanchored model, the relative effect parameter vector $\dd$ is of length $c$. This is because in the anchored models, $\dda$ is forced to be a vector of relative effects relative to treatment 1, and $d_{1,1} = 0$, so only the remaining $c-1$ relative effects need to be estimated. In contrast, in the unanchored model, the relative effects in $\dd$ are not restricted to be relative to any particular treatment. Therefore, it is necessary to estimate a relative effect for all $c$ components. In the unanchored model of R{\"u}cker et al. (2020),\citep{rucker2019network} estimation of $\dvec$ via weighted least squares and the Moore-Penrose generalized inverse automatically selects the anchor which gives the best fit to the data.

We also remark the following more flexible aspects of unanchored models:
\begin{itemize}
    \item The estimated anchor may not be a treatment implemented in the network. 
    \item The estimated anchor may be zero, corresponding to additivity on the absolute scale, rather than a relative scale.
\end{itemize}

We emphasize that anchored models impose the anchor treatment, which may or may not result in an additivity assumption which fits the data well. On the other hand, unanchored models estimate the anchor based on the best fit to the data. From this perspective, anchored models have the potential to be biased due to misspecifying the anchor. Due to a lack of discussion on this topic in the NMA literature, it seems unlikely that practitioners will have evidence for which treatment is the true anchor. Anchored models should be used with caution (see the simulation study in Section \ref{sec:sims}). In the following section, we provide two new Bayesian models which are unanchored.

\section{Novel Bayesian Unanchored CNMA Models}\label{sec:newmodels}

 We have defined anchored and unanchored models in the previous section, and shown that unanchored models are less prone to misspecification. Therefore, one may prefer to use CNMA models which are unanchored. Although the frequentist approach of R{\"u}cker et al. (2020) \citep{rucker2019network} is already an option for practitioners to implement an unanchored additive model, our Bayesian models will likely appeal to a wider audience given the popularity of Bayesian approaches. Additionally, the model of R{\"u}cker et al. (2020) \citep{rucker2019network} only accommodates contrast-level data. Thus, in Subsections \ref{sec:condat} and \ref{sec:armdat} we introduce novel unanchored Bayesian additive CNMA models to handle contrast-level and arm-level summary data, respectively. Table \ref{tab:models} contains a summary of the models included in this paper.

\subsection{Novel Bayesian Unanchored Model for Contrast-level Data} \label{sec:condat}

We introduce a random effects model which follows a similar formulation to R{\"u}cker et al. (2020) \citep{rucker2019network} in that it does not require the specification of an anchor for additivity and uses contrast-level data, however it uses a Bayesian framework. Additionally, in this model, we consider a more restrained set of contrasts than the model described in Subsection \ref{sec:rucker} which used all $\binom{a_i}{2}$ contrasts from study $i$.  Instead, we use as data only $a_i-1$ independent contrasts from each study $i$ by choosing a treatment $b_i$ in trial $i$ to be used as the common subtrahend when constructing the contrasts. The reason we take this approach is that using independent contrasts makes more sense from a Bayesian perspective (for example, see Dias et al. (2014) \citep{dias2014generalised}). Due to the different set of data being used compared to the model of R{\"u}cker et al. (2020),\citep{rucker2019network} our approach also handles correlation in multi-arm studies differently. While the approach by R{\"u}cker et al. (2020) \citep{rucker2019network} adjusts the standard errors, which in turn adjust the weights in the weighted least squares problem, we incorporate the correlation using covariance terms in the matrix of standard errors, which reflect the fact that each comparison in study $i$ is being taken relative to a common treatment. Nevertheless, the two different approaches to handling multi-arm studies should be equivalent.\citep{rucker2014reduce} Due to the way we handle multi-arm trials, some elements in our model have reduced dimension and will be indicated with superscript $*$.

Without loss of generality, assume the treatment employed in arm 1 of study $i$ is the treatment chosen to be the subtrahend in each contrast from that study. Let $\yyi^*$ be the $(a_i-1)\times1$ vector of independent contrasts relative to the treatment in arm 1 of study $i$ for $i=1,\dots,I$.  Let $SE_{ij}$ be the standard error of the contrast associated with the $j$th treatment arm compared to arm 1 in the $i$th study, and let $SE_{b_i}$ be the standard error of the raw response in arm 1 in study $i$ (the raw response to the treatment used as the common comparator in study $i$). Then we have
\begin{equation}
    \yyi^* \mid \mmu_i^*, \sigma^2 \sim \N(\mmu_i^*, \bigS_i^*),  i = 1, \dots, I,
\end{equation}
where
\begin{equation}
    \bigS_i^* =
    \begin{pmatrix}
    SE_{i2}^2 & \hdots & SE_{b_i}^2 \\
    \vdots & \ddots & \vdots \\
    SE_{b_i}^2 & \hdots & SE_{ia_i}^2
    \end{pmatrix}_{(a_i-1)\times(a_i-1)},
\end{equation}
noting that the diagonal elements of $\bigS_i^*$ represent variances of contrasts and the off-diagonal elements represent the variance of the responses in arm one of study $i$.
Further,
\begin{equation}
    \mmu_i^* = {\Bi}^* \Ai \dd + \epsss_i, \label{eq:contrastmu}
\end{equation}
where ${\Bi}^*$ is a $(a_i-1) \times a_i$ design matrix, where $U_{i,jk}^* = 1$ if the $k^\text{th}$ arm is the minuend of the $j^\text{th}$ contrast, $-1$ if the $k^\text{th}$ arm is the subtrahend in the $j^\text{th}$ contrast, and $0$ otherwise. The rows of ${\Bi}^*$ sum to zero and the first column of ${\Bi}^*$ (the common subtrahend) contains all $-1$'s. $\Ai$ is the same design matrix as introduced in Subsection \ref{sec:rucker}.
$\epsss_i$ represents between-study heterogeneity, and is distributed according to $\epsss_i \sim \N(\zero, \sigma^2\Sigma_i)$, where
\begin{equation}
\Sigma_i^* = \begin{pmatrix}
    1 & \hdots & 1/2 \\
    \vdots & \ddots & \vdots \\
    1/2 & \hdots & 1
    \end{pmatrix}_{(a_i-1)\times(a_i-1)}.
\end{equation}
Here, we also note that the inclusion of multicomponent treatments does not impact the compound symmetry of treatments. That is, the correlation between the heterogeneity terms of any two treatments in the same trial, regardless of if they are single or multicomponent, is 1/2.

To finish the Bayesian treatment of the model, we place vague priors on $\dvec$ and $\sigma$. Similarly to the model in Subsection \ref{sec:rucker}, in this model $\dvec$ represents a set of relative effects with an unknown reference point, corresponding to an unanchored model. We use priors similar to those used on the relative effects in the model described in Subsection \ref{sec:welton}, $\dvec \sim N(\zero, 1000\boldsymbol{I})$, and a uniform prior on the heterogeneity parameter $\sigma$ with a suitably large upper limit based on the scale of the data. 

\subsection{Novel Bayesian Unanchored Model for Arm-level Data}\label{sec:armdat}

In order to accommodate arm-level data, we adapt the model introduced in Subsection \ref{sec:condat}. We assume that the arm-level response consists of binomial counts of an outcome of interest for each arm of each study. However, it is important to note that the binomial distribution family and logit link functions we use can be modified to accommodate other types of outcomes, such as event counts or continuous outcomes, following Dias et al. (2014).\citep{dias2014generalised} Let the observed binomial counts in the $j^{\text{th}}$ arm of study $i$ be $r_{ij}$, and we assume $r_{ij} \mid n_{ij}, p_{ij}$ $\sim \Bin(n_{ij}, p_{ij})$, where $n_{ij}$ is the total number of subjects observed in arm $j$ of study $i$ and $p_{ij}$ is the probability of an outcome in the respective arm and study. Let $\probi = (p_{i1}, \dots, p_{ia_i})'$ be the vector of binomial probabilities in trial $i$. Then we model the binomial probabilities according to
\begin{equation}
    \logit(\probi) = \alpha_i\one + \Ai \dd + \epsi, \hspace{0.5cm}  i = 1, \dots, I, \label{eq:armmodel}
\end{equation}
where
\begin{equation}
    \epsi \sim \N(\zero, \sigma^2\Sigma_i), \notag
\end{equation}
$\logit(\probi)$ is a vector of the log-odds of outcomes in the arms of study $i$, $\alpha_i$ is the baseline response in study $i$ to the unknown anchor, $\sigma$ represents the degree of between-study heterogeneity, $\Sigma_i$ is given in equation \eqref{eq:epsidist}, the vector $\one$ is an $a_i\times 1$ vector of ones, and $\Ai$ is a design matrix with dimension $a_i \times c$ as defined in Subsection \ref{sec:rucker}. Additionally, $\dd$ is a $c \times 1$ vector of relative effects, relative to the unknown anchor. For priors, we employ the same as those used in Subsection \ref{sec:welton}.

The model proposed herein is simply an equivalent to the model of the previous section for data presented at the arm level. A proof of this assertion is presented in Appendix \ref{sec:proof}. We show that equation \eqref{eq:contrastmu} is a contrast-based equivalent to equation \eqref{eq:armmodel}, and the two models are equivalent, with the exception that the contrast-based method requires a normality assumption when calculating the standard errors in the matrix $\bigS_i^*$.

\begin{table}[htbp]
  \centering
  \caption{Summary table of the four CNMA models described in this paper, labeled by the subsection in which they were introduced (2.1, 2.2, 4.1 or 4.2.) Note that the definition of $\Sigma_i$ in the Subsection 2.1 model is altered if trial $i$ includes the reference treatment.}

    \begin{tabular}{rcc}\hline
          & \multicolumn{2}{c}{Existing models} \\
          & 2.1 (arm, anchored)   & 2.2 (contrast, unanchored) \\\hline
    Data  & $r_{ij} \sim \text{Binomial}(n_{ij}, p_{ij})$ & $\yy_i  \sim N(\mmu_i, \Omg_i)$ \\
    Treatment effects & $\logit(\probi) = 
    \alpha_i\one + \Aistar \dda + \epsi$ & $\mmu_i = \Bi\Ai\dd$ \\
    Heterogeneity & $\epsi \sim \N(\zero, \sigma^2\Sigma_i)$ & $\hat \sigma^2$, embedded in $\Omg_i$ \\\hline
          & \multicolumn{2}{c}{New models} \\
          &  4.2 (arm, unanchored)  & 4.1 (contrast, unanchored) \\\hline
    Data  & $r_{ij} \sim \text{Binomial}(n_{ij}, p_{ij})$  & $\yy_i^* \mid \mmu_i^*, \sigma^2 \sim N(\mmu_i^*, \bigS_i^*)$ \\
    Treatment effects &  $\logit(\probi) = \alpha_i\one + \Ai \dd + \epsi$     &  $\mmu_i^* = {\Bi}^* \Ai\dd + \epsi^*$ \\
    Heterogeneity &  $\epsi \sim \N(\zero, \sigma^2\Sigma_i)$   & $\epsi^* \sim \N(\zero, \sigma^2\Sigma_i^*)$   \\
    \end{tabular}%
  \label{tab:models}%
\end{table}%

\section{Simulation Study}\label{sec:sims}

We have clarified the two approaches to assuming additivity made in existing CNMA models, and introduced two Bayesian models which avoid the the need to specify an anchor a priori. The simulations in this section are designed to illustrate the biases which can arise from misspecifying the anchor in an anchored model and demonstrate the performance of the new unanchored models in Section \ref{sec:newmodels}. To this end, we consider eight different simulation scenarios (described in Subsection \ref{subsec:scenarios}), and simulate 1000 arm-level datasets from each scenario using the model presented in Subsection \ref{sec:welton} with the anchor treatment specific to the scenario. We choose this model to generate data because it allows for the generation of arm-level data and the specification of the true anchor in additivity. 

Each dataset is analysed using the four models covered in this paper: Bayesian anchored CNMA model with arm-level data (referred to as Bayes Arm Anchored), Frequentist unanchored CNMA model with contrast-level data (Freq Contrast Unanchored), Bayesian unanchored CNMA model with contrast-level data (Bayes Contrast Unanchored), and Bayesian unanchored CNMA model with arm-level data (Bayes Arm Unanchored). For models Freq Contrast Unanchored and Bayes Contrast Unanchored, the arm-level data were converted to contrast-level data using the procedure described in Appendix \ref{sec:dgm}. 

The priors specified for the Bayesian models were $\dd$ or $\dda \N(\zero, 1000 \bm{I})$ for the unanchored and anchored models respectively, $\sigma \sim \text{Unif}(0, 2)$, and $\alpha \sim \N(0, 1000)$ for the arm-based models.

Bayes Arm Anchored requires specifying an anchor treatment. In the simulations, the placebo is always specified as the anchor in the analysis for this model, and is indicated in brackets after the model name. In some scenarios the placebo is the true anchor and in others it is misspecified. This is indicated by including the data-generating, true anchor in the plot labels. When the analysis anchor is different from the data-generating anchor, the Bayes Arm Anchor model is misspecified.

\subsection{Scenarios}\label{subsec:scenarios}

\begin{figure}
    \centering
    \includegraphics[scale = 0.65]{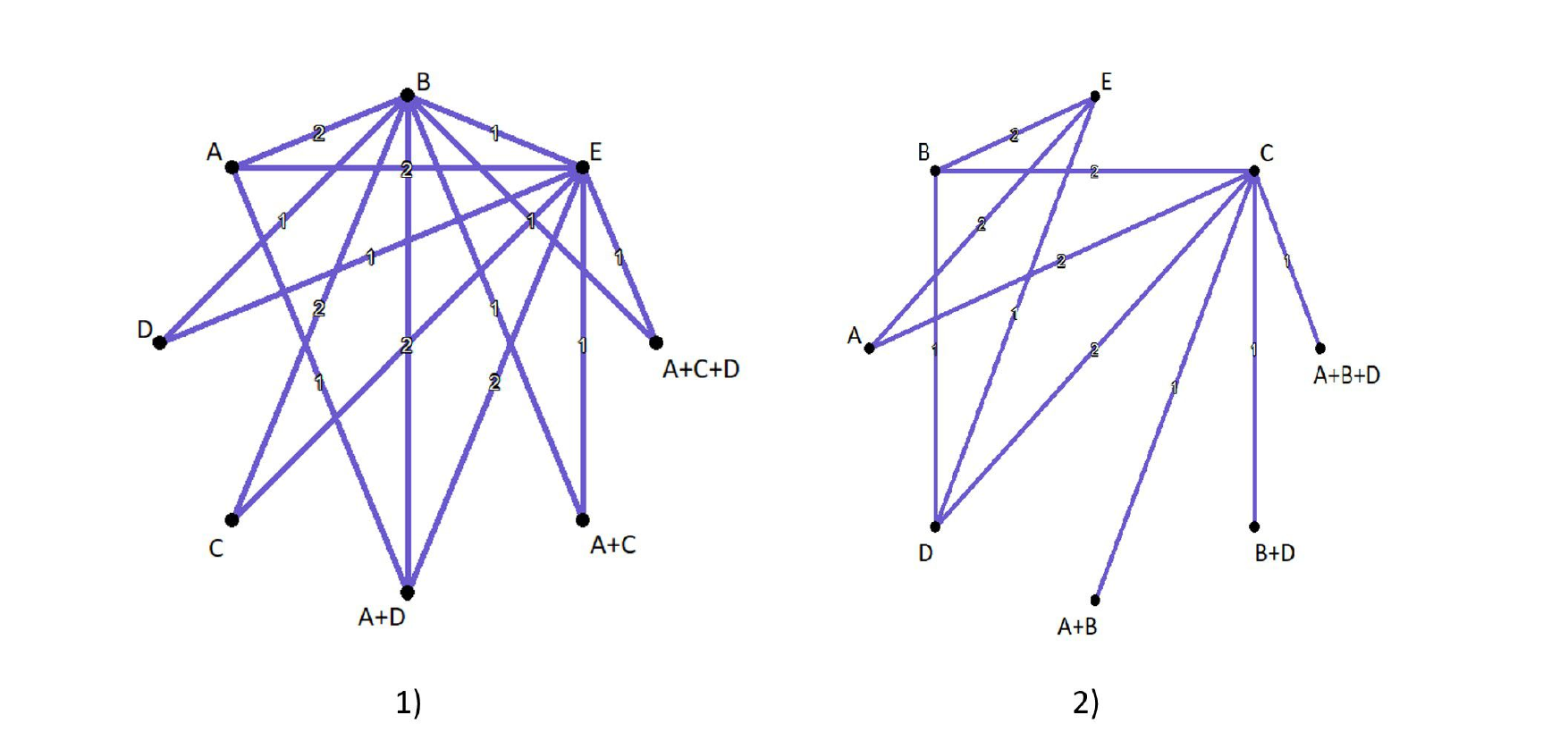}
    \caption{Network plots of the two scenarios considered in the simulation study. Nodes represent treatments and edges represent trials comparing those treatments, with the number of such trials overlaid. On the left, labelled 1), the ``worst-case scenario" for misspecifying the anchor in additivity. On the right, labelled 2), the more moderate scenario.}
    \label{fig:network}
\end{figure}

Two different network structures were used to reflect different simulation scenarios, and are presented in Figure \ref{fig:network}. Both network structures include five single-component interventions A, B, C, D, and E, and three multicomponent interventions. The treatments are labelled alphabetically according to decreasing effectiveness in each network structure, assuming a large response is good. All studies have two arms. In each network, additivity was considered relative to two different anchor treatments. In each of the two networks and for each of the two additivity assumptions, two different amounts of data were considered: 50 and 500 patients per arm.


In the first network (referred to as network 1), the relative effects are set so that the worst treatment (treatment E) is much less effective than the other four treatments. The value of the heterogeneity ($\sigma$) is set relatively low. The parameter values used in simulating data for this network are presented in the first three columns of Table \ref{tab:networkparams}. We consider two different ways to simulate the multicomponent data: one where the true anchor is the treatment E; the other specifying the true anchor to be a different treatment, B, which has a similar effect size to the remaining treatments A, C and D. The additivity assumption implied by using E as the anchor treatment is quite different from the additivity assumption implied by using treatment B as the anchor. This can be seen in the multicomponent treatment effects in Table \ref{tab:networkparams} and is due to the fact that treatments E and B have quite different effect sizes. From this table, we can see that misspecifying the anchor as E when its true value is B could give poor results - our assumption (that the true anchor is E) expects the multicomponent treatment effects shown in the third column of table \ref{tab:networkparams}, but the data-generating mechanism reflects the multicomponent treatment effects in the second column. Therefore, network 1 represents a scenario where misspecifying the anchor treatment has a strong impact on model fit. This could correspond to a scenario where a set of trials compared several very effective drugs to a placebo treatment (which is not very effective), and the true anchor is one of the effective drugs but the data is analysed assuming the placebo is the anchor.

\begin{table}[htb]
    \centering
    \caption{Parameter values used in the simulation for Network 1 and Network 2. Values of the parameters under different anchors are also given. The value of the single component relative effects and the heterogeneity $\sigma$ are the same for different anchors. The multicomponent relative effects and the baseline responses are different for different anchor treatments.}
    \begin{tabular}{c|c|c|c|c|c}
    \multicolumn{3}{c|}{Network 1} & \multicolumn{3}{c}{Network 2} \\
    \hline \hline
    Parameter & Anchor B & Anchor E & Parameter & Anchor C & Anchor E \\
    \hline
    $d_{E,A}$ & 1.2 & 1.2 & $d_{C,A}$ & 0.70 & 0.70 \\
    $d_{E,B}$ & 0.9 & 0.9 & $d_{C,B}$ & 0.35 & 0.35\\
    $d_{E,C}$ & 0.8 & 0.8 & $d_{C,C}$ & 0.00 & 0.00\\
    $d_{E,D}$ & 0.7 & 0.7 & $d_{C,D}$ & -0.20 & -0.20 \\
    $d_{E,E}$ & 0.0 & 0.0 & $d_{C,E}$ & -0.50 & -0.50\\
    \hline
    $d_{E,A+C}$ & 1.1 & 2.0 & $d_{C,A+B}$ & 1.05 & 1.55\\
    $d_{E,A+D}$ & 1.0 & 1.9 & $d_{C,B+D}$ & 0.15 & 0.65\\
    $d_{E,A+C+D}$ & 0.9 & 2.7 & $d_{C,A+B+D}$ & 0.90 & 1.85\\
    \hline 
    $\alpha$ & 0.05 & -0.85 & $\alpha$ & -0.60 & -1.10 \\
    \hline
    $\sigma$ & 0.1 & 0.1 & $\sigma$ & 0.40 & 0.40
    \end{tabular}
    \label{tab:networkparams}
\end{table}

The second network (network 2) was designed to represent a more moderate scenario in terms of the consequences of misspecifying the anchor for additivity. The parameter values used in simulating data for this network are presented in the last three columns of Table \ref{tab:networkparams}. The heterogeneity in the network ($\sigma$) was increased compared to network 1. We consider two different ways to simulate the multicomponent data: one specifying the true anchor to be the moderately effective treatment C; the other specifying the true anchor to be the worst treatment, E. The absolute relative effect of those two treatments is smaller (0.5) compared to the absolute difference of treatments E and B in network 1 (0.9) and there are fewer direct and indirect studies providing data on the relative effect. Network 2 represents a less severe scenario in terms of misspecifying the anchor treatment because the two additivity assumptions in network 2 give more similar multicomponent treatment effects than the two additivity assumptions used in network 1. For details on how the values in Table \ref{tab:networkparams} were used to simulate the data, see Appendix \ref{sec:pars}. The resulting relative effects of multicomponent treatments are also presented in Table \ref{tab:networkparams}.

\subsection{Performance Measures}\label{subsec:perf}

For the Bayesian models, the posterior mean was used as the point estimate. To quantify the bias in each model, we calculate the difference between the estimated value and the true value used in data generation for each parameter. Coverage probability was calculated for 95\% confidence and credible intervals for the frequentist and Bayesian models respectively. Additionally, length of the confidence and credible intervals was calculated. 

To assess the ability of the models to give accurate treatment rankings, we computed the average surface under the cumulative ranking (SUCRA) curve for the Bayesian models, and average P-score for the frequentist model. The SUCRA value and P-score summarize the information about a treatment's probability of being best.\citep{rucker2015ranking, salanti2011graphical} SUCRA values were calculated using the dmetar R package and P-scores were calculated using the netmeta R package. 

Methods to assess model fit in terms of additivity and choice of anchor are of interest. We investigate the use of DIC to assess and compare model fit between anchored and unanchored models. In particular, we compare the anchored arm-level model \ref{sec:welton} to the unanchored arm-level model \ref{sec:armdat} using the DIC. Let $\Delta = DIC(\text{Model 4.2}) - DIC(\text{Model 2.1})$. Then negative values of $\Delta$ indicate better fit for the unanchored model compared to the anchored model. Differences in DIC of two or more are generally considered meaningful.\citep{spiegelhalter2002bayesian}

\begin{figure}
    \centering
    \makebox[\textwidth]{\includegraphics[width = 7.5in]{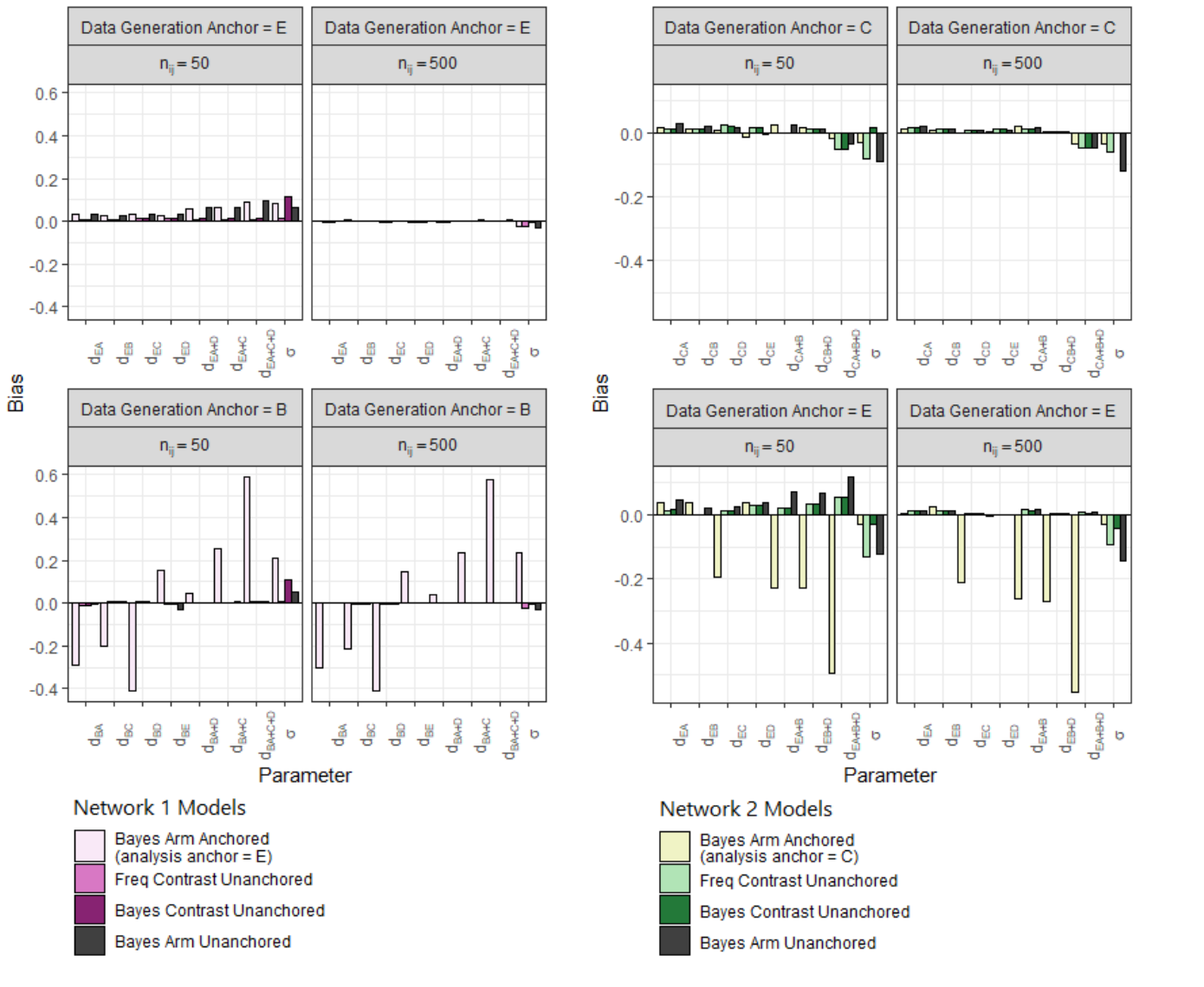}}
    \caption{Bias of estimates of relative effect parameters and heterogeneity for Network 1 (left, pink) and Network 2 (right, green). The relative effect parameters shown correspond to treatment effects relative to whichever treatment was used in the data generation for that scenario, for example in the top left, the anchor used to generate data was E so the relative effects shown are $d_{E,k}$, which can also be seen in the x-axis labels.}
    \label{fig:biasresults}
\end{figure}

\begin{figure}
    \centering
    \makebox[\textwidth]{\includegraphics[width = 7.5in]{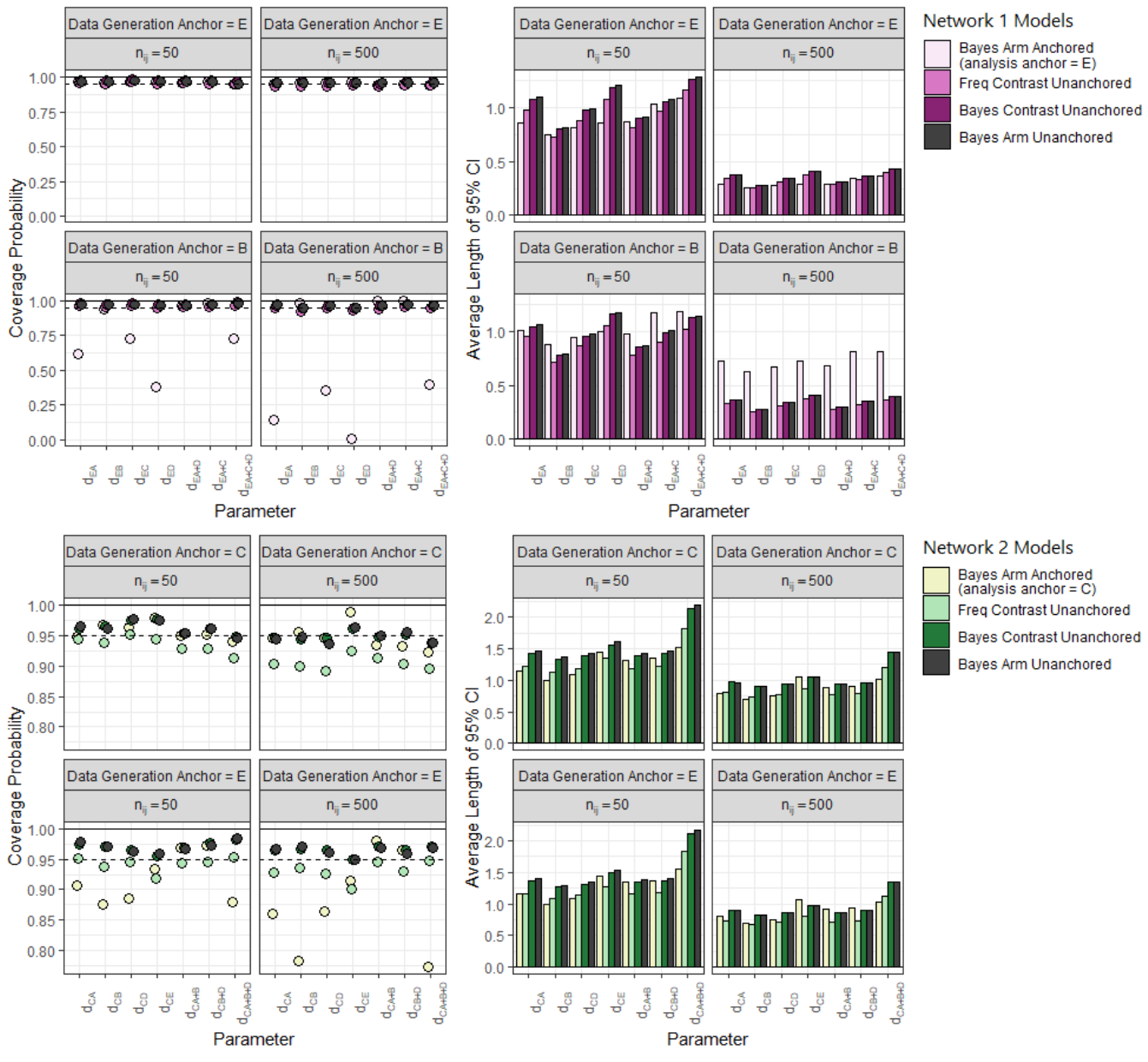}}
    \caption{Coverage probability and average length of 95\% credible or confidence intervals for the relative effect parameters monitored in the simulation for network 1 (top row, pink) and network 2 (bottom row, green). The relative effect parameters shown correspond to the relative effects monitored in the Bayesian analyses, corresponding to $d_{E,k}$ for network 1 and $d_{C,k}$ for network 2, which can also be seen in the x-axis labels.}
    \label{fig:ciresults}
\end{figure}




\subsection{Network 1 Results}

The bias for relative effect parameters and heterogeneity is shown for network 1 on the left of Figure \ref{fig:biasresults}. When the anchor used in data generation is E, the bias for the relative effects is small for all models and sample sizes. When the treatment B is used as the anchor for data generation, the bias for the three unanchored models remains small, however the Bayes Arm Anchored (analysis anchor = E) model has much larger magnitude biases, regardless of the sample size. The bias for $\sigma$ decreases as the sample size increases for the Bayesian models, but it increases for the frequentist model. Overall, the bias is always relatively low for the unanchored models, but when the anchor is misspecified, the anchored model shows severe bias. Here we also note that due to the assumption of consistency, if the set of relative effect estimates with one treatment as the comparator are unbiased, then the sets of relative effect estimates with any other treatment as the comparator will also be unbiased. Similarly, if the set of relative effect estimates with one treatment as the comparator are biased, then the sets of relative effect estimates with any other treatment as the comparator will also likely be biased.

The coverage probability and average length of 95\% credible/confidence intervals for network 1 is shown on the top row of Figure \ref{fig:ciresults}. For the unanchored models, the coverage probability is close to the nominal level of 0.95 for all relative effect parameters, sample sizes, and data generation anchors. However, when the analysis anchor does not match the data generation anchor, the anchored model performs poorly. In that scenario, $d_{E,A}$, $d_{E,C}$, $d_{E,D}$ and $d_{E,A+C+D}$ all have coverage probabilities well below the nominal level of 0.95 which decrease as the sample size is increased, falling to zero for $d_{E,D}$. Interestingly, the single component treatments whose relative effects have poor coverage are the components involved in the multicomponent treatments.

Turning to the average lengths of the credible intervals, when the data generation anchor is E, models Bayes Arm Anchored (analysis anchor = E) and Freq Contrast Unanchored have the shortest credible intervals and the lengths decrease for the larger sample size. When the data generation anchor is B, Bayes Arm Anchored (analysis anchor = E) model tends to have longer credible intervals, particularly for the larger sample size compared to the other models.

\begin{figure}
    \centering
    \makebox[\textwidth]{\includegraphics[width = 7.5in]{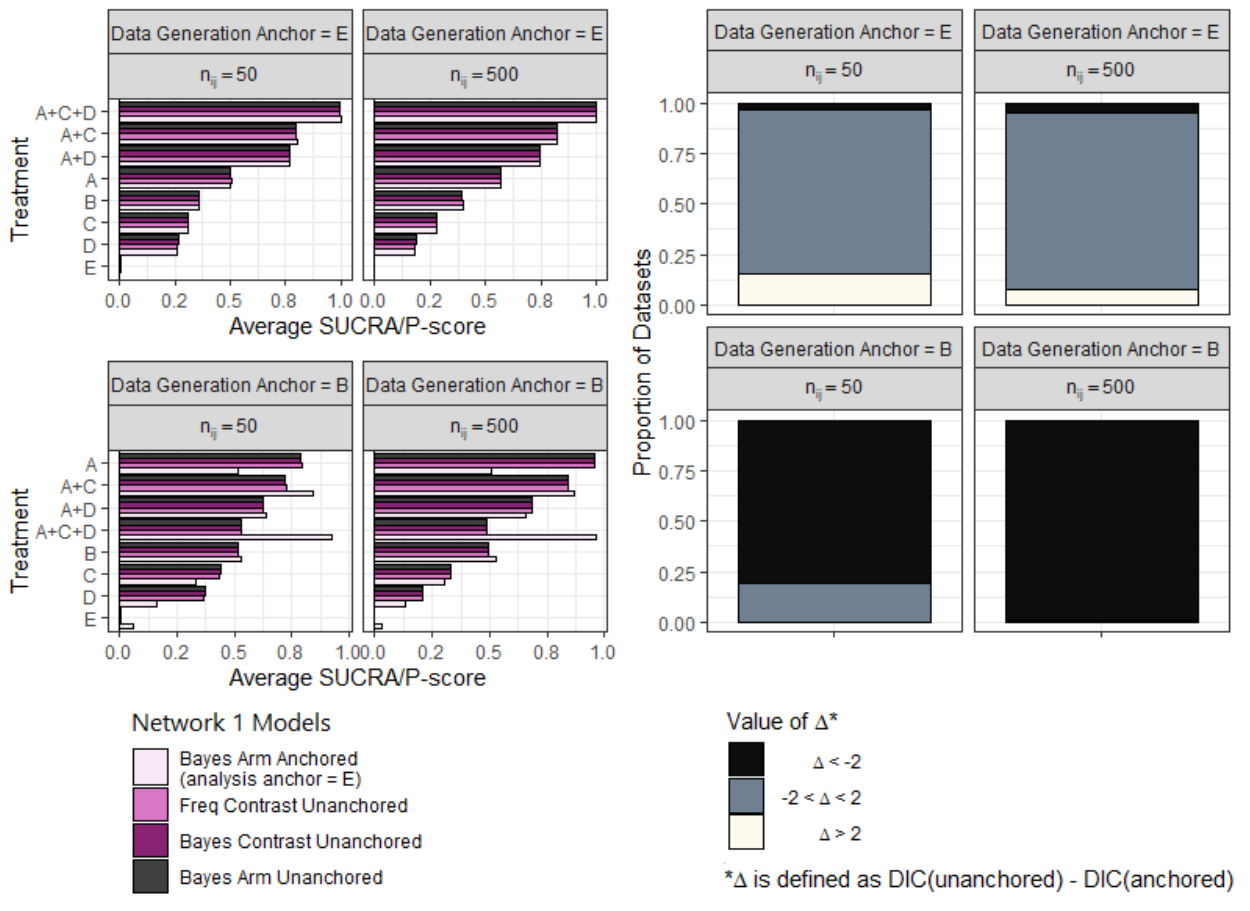}}
    \caption{Network 1 average rankings based on SUCRA/P-scores and percentage of datasets where the value of $\Delta = \text{DIC(Model 3.2)} - \text{DIC(Model 2.1)}$ is less than -2, between -2 and 2, and greater than 2. Negative values of $\Delta$ indicate that the unanchored model has a better fit. Differences of DIC greater than 2 are considered meaningful.\citep{spiegelhalter2002bayesian}}
    \label{fig:psoc2}
\end{figure}

The average SUCRA or P-score value for each treatment in the different scenarios is shown on the left of Figure \ref{fig:psoc2}. The order of the treatments on the Y-axis indicates their true treatment rankings, where the treatment with the largest effect is at the top and the smallest effect is at the bottom. The true order changes depending on which treatment was used as the anchor in generating the data. When treatment E is the true anchor, all four models do a good job on average of choosing the rankings based on the SUCRA and P-scores. However, when treatment B is the true anchor, the Bayes Arm Anchored (analysis anchor = E) model results in average SUCRA scores which reflect a different ranking of the treatments than the truth. In particular, the treatment ranking implied by the average SUCRA scores from this model is the same as the treatment ranking which is true when E is the true anchor.

A breakdown of the values of $\Delta$ for each scenario comparing the DIC of the unanchored and anchored Bayesian arm-level models is given in the right panel of Figure \ref{fig:psoc2}. When the true anchor is E, in the majority of the models, neither the anchored or unanchored models have a meaningfully better fit than the other model, for both sample sizes. When the true anchor is B, the unanchored model tends to be preferred by the DIC the majority of the time. The Bayes Arm Anchored (analysis anchor = E) model is recommended only 0.03\% of the time when $\nij = 50$ and is never recommended when $\nij = 500$.

\subsection{Network 2 Results: Placebo and Waitlist Network}

The bias of estimates in network 2 is shown on the right side of Figure \ref{fig:biasresults}. The bias for the relative effect parameters shows a similar pattern to network 2. In particular, the anchored model, Bayes Arm Anchored (analysis anchor = C), shows large biases when the analysis anchor is misspecified. For both data generation anchors, the magnitudes of the bias for the heterogeneity are larger for the Freq Contrast Unanchored and Bayes Arm Unanchored models compared to the other two models.

The coverage probability and average length of 95\% credible/confidence intervals for relative effects are shown in the bottom row of Figure \ref{fig:ciresults}. When the data generation anchor is C, the coverage probabilities are close to the nominal level for all models and sample sizes, although the Freq Contrast Unanchored model tends to fall slightly below 0.95 and the Bayesian models slightly above 0.95. When the data generation anchor is E, the Bayes Arm Anchored (analysis anchor = C) model has coverage probabilities which fall below the nominal level. The lengths of credible and confidence intervals tend to be shortest for models Bayes Arm Anchored (analysis anchor = C) and Freq Contrast Unanchored when the data generation anchor is C. As the sample size increases from 50 to 500, the average length of the credible intervals decreases for all models. A similar pattern is seen when the data generation anchor is E.

\begin{figure}
    \centering
    \makebox[\textwidth]{\includegraphics[width = 7.5in]{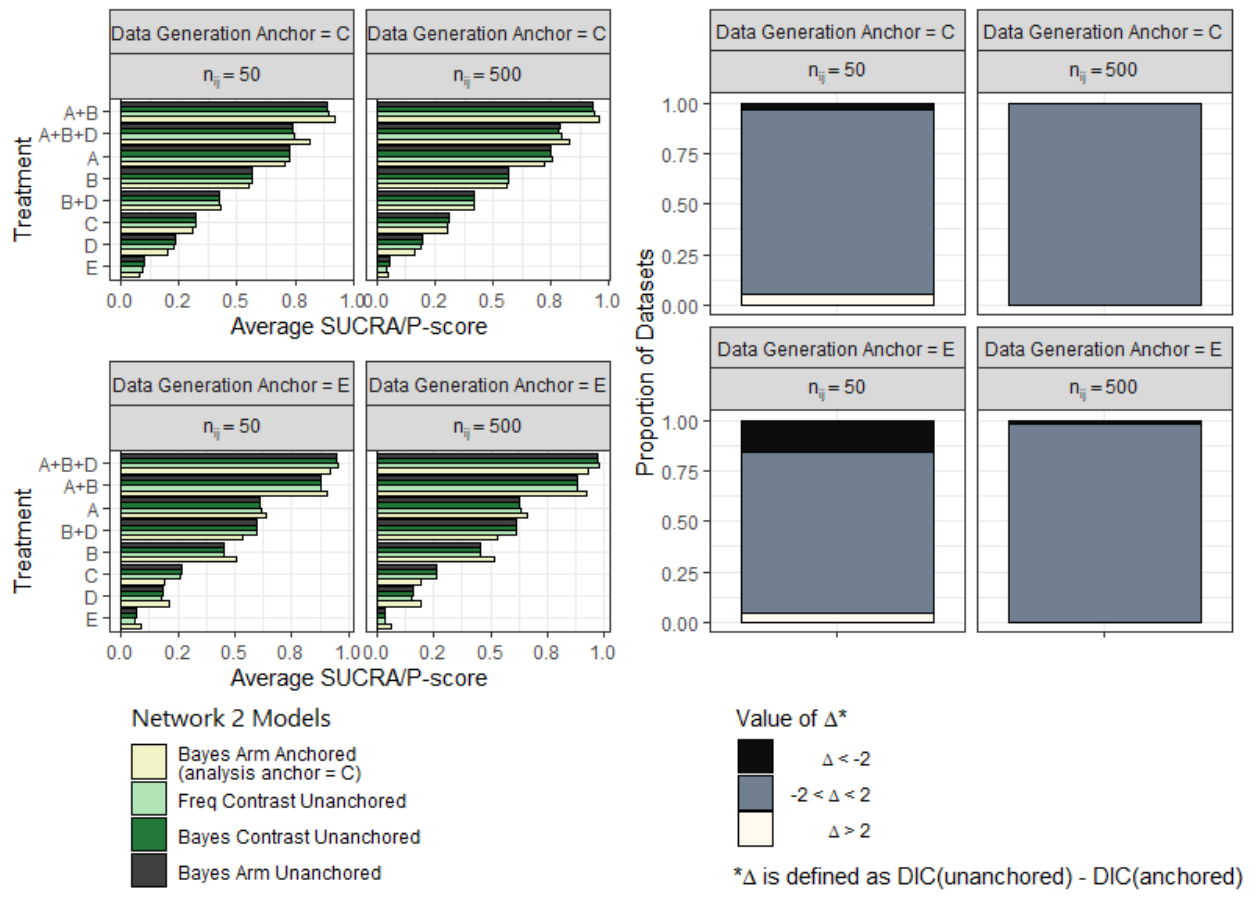}}
    \caption{Network 2 average rankings based on SUCRA/P-scores and percentage of datasets where the value of $\Delta = \text{DIC(Model 4.2)} - \text{DIC(Model 2.1)}$ is less -2, between -2 and 2, and greater than 2. Negative values of $\Delta$ indicate that the unanchored model has a better fit. Typically, differences of DIC greater than 2 are considered meaningful.\citep{spiegelhalter2002bayesian}}
    \label{fig:pwl2}
\end{figure}

The left panel of Figure \ref{fig:pwl2} shows the average SUCRA or P-score values for each treatment in the different scenarios. Similar results are seen for both sample sizes for both anchor scenarios. When the true anchor is C, all four models give average SUCRA/P-scores that correspond to the true treatment ranking. The Bayes Arm Anchored (analysis anchor = C) tends to give higher average SUCRA scores to the best treatments ($A+B$, $A+B+D$) and lower scores to the worse treatments ($C$, $D$, $E$) than the other three models. When the true anchor is E, all models give average SUCRA/P-scores corresponding to the correct treatment ranking, with the exception of Bayes Arm Anchored (analysis anchor = C), which incorrectly ranks $D$ slightly higher than $C$ on average. The right panel of Figure \ref{fig:pwl2} shows a breakdown of the values of $\Delta$ for the different anchor and sample size scenarios for network 2. In all scenarios, the most common outcome is that the DIC does not recommend one model over the other. 

\section{Application} \label{sec:casestudy}

In this section, we use the two novel CNMA models introduced in Section \ref{sec:newmodels} to analyse the dataset utilised by Welton et al. (2009) \citep{welton2009mixed} which regards clinical trials of psychological interventions for adults with coronary heart disease (CHD). We present the design matrices for the novel models for one trial to illustrate their construction.
We also analyse the data using the existing additive CNMA models of Welton et al. (2009) \citep{welton2009mixed} and R{\"u}cker et al. (2020) \citep{rucker2019network} for the sake of comparison. Finally, we compare the model fit of the two Bayesian arm-level models using the DIC, similarly to the simulation study.

We focus on the studies which report all-cause mortality as one of the outcomes. This is a binary outcome. The dataset contains 36 trials which report all-cause mortality. Welton et al. (2009) \citep{welton2009mixed} classified the psychological interventions employed in the clinical trials as being educational (Edu), behavioural (Beh), cognitive (Cog), relaxation (Rel), supportive (Sup), or some multicomponent combination of these. In all the trials included, the psychological interventions were compared to a control arm which employed usual care. Several of the studies recorded zero events.

For illustration purposes, we show the design matrices for the multiarm trial numbered 23 for models \ref{sec:condat} and \ref{sec:armdat}. This trial has three treatment arms. In the first arm, a psychological intervention composed of educational, cognitive, and relaxation components was investigated. In the second arm, a psychological intervention composed of educational and relaxation components was used. The third arm used usual care only. For the contrast-based unanchored model introduced in Subsection \ref{sec:condat}, we require two design matrices: $\BB_i^*$ and $\Ai$. $\BB_i^*$ forms the contrasts between the arms, where the columns represent arms and the rows represent contrasts. Recall that in this model, $a_i-1$ contrasts are used. Assuming that the two contrasts used are arm two versus arm one and arm three versus arm one, $\BB_i^*$ for this trial is given by
\begin{equation}
\BB_i^* = \begin{pmatrix}
-1 & 1 & 0 \\
-1 & 0 & 1
\end{pmatrix}.
\end{equation}
The design matrix $\Ai$ shows the component structure of the arms, and is given by 
\begin{equation}
\Ai = \begin{pmatrix}
0 & 1 & 0 & 1 & 1 & 0 \\
0 & 1 & 0 & 0 & 1 & 0 \\
1 & 0 & 0 & 0 & 0 & 0 \\
\end{pmatrix}.
\end{equation}
In this matrix, the columns correspond to the components Usual, Edu, Beh, Cog, Rel, and Sup. Model \ref{sec:armdat} uses the same $\Ai$ as model \ref{sec:condat}.

\subsection{Results}

Random-effects additive CNMA models were fit to the CHD dataset to compare and contrast the results of the four CNMA models. The Bayesian models were fit to the data in R using JAGS for MCMC sampling via the runjags R package.\citep{denwood2016runjags} The models ran two parallel chains and at least $10,000$ burn-in iterations followed by at least $40,000$ samples from the posterior. Convergence was assessed by inspecting trace plots. The R package netmeta was used to fit a random-effects additive CNMA model using the model introduced in Subsection \ref{sec:rucker}.\citep{netmeta} 

For the purposes of comparing the anchored and unanchored model fit, DIC was calculated for the Bayes Arm Anchored and Bayes Arm Unanchored model using JAGS. Although it is typically recommended that the DIC be calculated manually for hierarchical models, no leverage can be manually calculated for study arms which experience zero events, and thus the DIC cannot be calculated manually for this data example.\citep{dias2018network} The DIC calculated automatically by JAGS is thus reported.

\begin{figure}
    \centering
    \includegraphics[scale = 0.6]{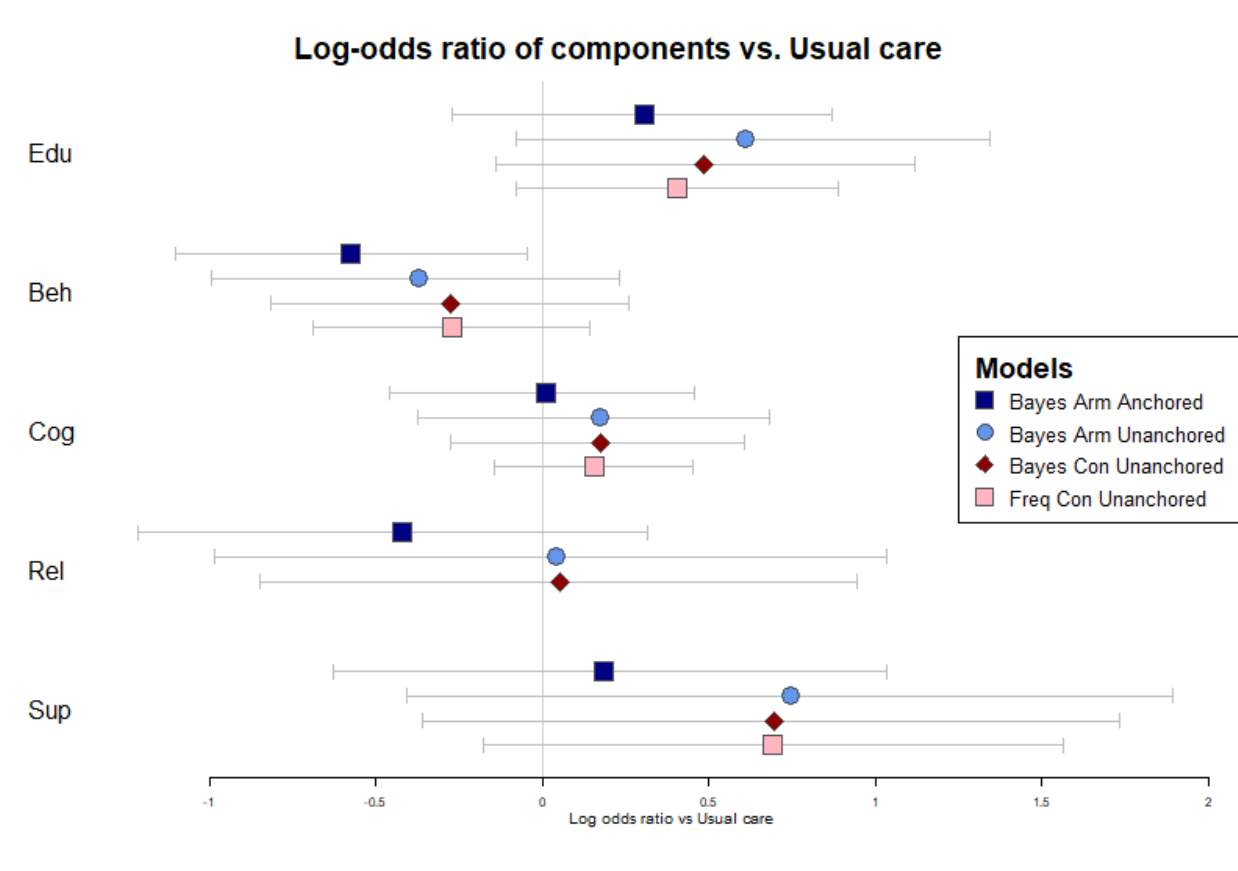}
    \caption{Log-odds ratios of the five psychological components versus usual care and 95\% confidence/credible intervals for the four additive CNMA models. Note that a point estimate and confidence interval is missing from the frequentist for the relaxation component due to this not being readily available from the R package netmeta.}
    \label{fig:casestudy}
\end{figure}

The results for the relative effects of the psychological components compared to usual care are shown in Figure \ref{fig:casestudy} in a forest plot. We note that due to the way the data was coded in the original model used by Welton et al. 2009,\citep{welton2009mixed} the multi-arm correction was not correctly applied to trials with three arms. By recoding the data so that usual care is implemented in the first arm of each study, this issue is corrected. Our results presented here therefore differ slightly from those in the original paper of Welton et al. 2009.\citep{welton2009mixed} The most important finding from this plot is that while all three unanchored models find no statistically significant differences between psychological interventions and usual care as the credible/confidence intervals include zero, the anchored model finds that behavioural interventions significantly reduce all-cause mortality (credible interval does not include 0). Additionally, the point estimates from the anchored model tend to be shifted downwards compared to those from the unanchored models, while those from the unanchored models are closer to each other. 

Although the conclusions from all three unanchored models are similar, some differences remain. Firstly, the frequentist model gives narrower confidence intervals than the Bayesian models' credible intervals. This could be due to the fact that the Bayesian models account for the uncertainty in the estimation of heterogeneity while the frequentist model treats the estimated heterogeneity parameter as fixed and known. Secondly, the Bayesian arm-level model \ref{sec:armdat} gives longer credible intervals than either of the other unanchored models and gives slightly different point estimates than the other unanchored models, particularly for educational and behavioural components. One possible explanation for this is that the contrast-based models make a normal approximation in calculating the standard errors of the contrasts, whereas the arm-level methods do not. This could result in loss of information in the data, particularly for studies for whom the normal approximation is poor, such as those with small sample sizes.

Overall, all three unanchored models give similar results and we note that the anchored model finds a significant effect of behavioural components whereas the unanchored models do not find any significant effects. The findings from this application support our assertion that specifying an anchor a priori imposes a restriction on the model.

We can compare the DIC for the two Bayesian models which use arm-level data. The anchored and unanchored models had a DIC of 363.4 and 363.5 respectively. From this, there is no evidence that one model has a better fit over the other.

\section{Discussion}\label{sec:disc}

In this paper, we introduced the notion of the anchor treatment and anchored and unanchored CNMA models. We introduced unified notation and clarified the nature of the additivity assumption in two existing CNMA models. We also presented arguments for why unanchored models are more robust than anchored models. We provided two novel Bayesian unanchored additive CNMA models to capitalize on the advantages of a Bayesian approach. A simulation study was completed to demonstrate the performance of the unanchored and anchored additive CNMA models under different scenarios in terms of relative effects, true anchor, and sample sizes. A case study was presented where all additive CNMA models were demonstrated on a real dataset. 

Anchored models such as Welton et al. (2009) \citep{welton2009mixed}'s can lead to serious problems if the anchor is misspecified. In the simulation study, we found that if the analysis anchor does not match the true anchor used to generate data, important biases can be introduced and coverage probability can be poor. The biases introduced can also lead to inaccurate treatment rankings based on SUCRA values. In these situations, unanchored models such as that of R{\"u}cker et al. (2020) \citep{rucker2019network} and the Bayesian models introduced in this paper are preferable. However, in the simulation study when the analysis anchor was correctly specified, the anchored model can present narrower CIs than the unanchored models. Thus, practitioners may still wish to use an anchored model if they are confident in their choice of the anchor treatment. 

Given that anchored models can provide benefits when the analysis anchor agrees with the data generating mechanism, but can lead to serious problems if it is misspecified, a method to test the suitability of a specific treatment as the anchor in the anchored model is desirable.
Welton et al. (2009) \citep{welton2009mixed} use the DIC to choose between standard NMA and CNMA models. 
In our simulation study, we explored the use of DIC for detecting an unsuitable anchor by comparing the anchored model to an unanchored model. However, we found that in network 2, despite notable biases and poor coverage probability of the misspecified anchor model, the DIC did not reliably recommend the unanchored model over the anchored model. A more sophisticated approach to testing the suitability of the analysis anchor for Bayesian anchored models is needed. Turning to frequentist methods, R{\"u}cker et al. (2020) \citep{rucker2019network} provided a likelihood ratio test for comparing the additive unanchored CNMA model and the standard NMA model. No frequentist anchored CNMA model has been developed at this time, so no tests are available to assess the anchor choice if frequentist models are used. Overall, development of methods to assess the suitability of the chosen anchor in anchored additive CNMA models are needed.

In this paper, we restricted our investigation to additive CNMA models. The additivity assumptions made in the two existing CNMA models discussed in this paper can be relaxed by including interaction terms.\citep{welton2009mixed,rucker2019network} The new Bayesian models introduced in Section \ref{sec:newmodels} could be extended this way as well. In addition, the Bayesian models could be applied to disconnected networks as described in R{\"u}cker et al. (2020) \citep{rucker2019network} and demonstrated in R{\"u}cker et al. (2020).\citep{rucker2020component} Another potential benefit of the Bayesian CNMA framework is that the additivity assumption could be relaxed by introducing structure in the priors for the relative effects of multicomponent treatments. This approach would not require interaction parameters to be estimated, which is attractive in situations where data are sparse, such as in disconnected networks. These present some important avenues for future work.

In summary, we identified a key difference in the additivity assumptions made in existing CNMA models and introduced the terminology of anchored and unanchored models. We presented two new Bayesian CNMA models which do not require the anchor to be specified and are thus unanchored. In the simulation study, we demonstrated the risks associated with using an anchored additive CNMA model. More work is needed to develop methods for assessing the anchor used in anchored CNMA models. The Bayesian unanchored CNMA models are robust and have the potential to be extended to relax additivity and be applied to disconnected networks. 


\section*{Acknowledgements}

The authors gratefully acknowledge the comments from reviewers and the Associate Editor, which improved the clarity of the paper.
The simulations in this research were enabled in part by the facilities of the Shared Hierarchical Academic Research Computing Network (SHARCNET:www.sharcnet.ca) and Compute/Calcul Canada (www.computecanada.ca).

\subsection*{Funding Acknowledgement}

A.W. gratefully acknowledges the support from the Ontario Graduate Scholarship program. A.B. gratefully acknowledges the support of the Natural Sciences and Engineering Research Council of Canada (RGPIN-2019-04404). 

\subsection*{Conflict of Interest}

Since 2020, A.W. has received consulting fees for work unrelated to this project from Covalence Research Ltd. for software developments related to network meta-analysis.

From 2017 to 2019, A.B. has received consulting fees for work unrelated to this project from ICON plc. and Lighthouse Outcomes Ltd. for methodological advice related to network meta-analysis.

\section*{Data Availability Statement}

The data and code used to produce the results in the application section are openly available in a GitHub repository at \url{https://github.com/augustinewigle/unanchored-cnma}. These data were derived from the following resource in the public domain: \url{http://www.bristol.ac.uk/population-health-sciences/centres/cresyda/mpes/code/}. The code used to obtain the results of the simulation study are available from the corresponding author upon reasonable request.

\appendix

\section{Proof of Uniqueness of Anchor Treatment}\label{sec:unique}

Suppose $X$, $Y$, and $Z$ are distinct treatments in an arbitrary network. Let $|X|>1$ denote the number of components in multi-component treatment $X$. Assume that the additivity equation holds for treatment $Y$ as the anchor and for treatment $Z$ as the anchor simultaneously, that is,
\begin{equation}
    d_{Y,X} = \sum_{c\in X} d_{Y,c} \label{yx}
\end{equation}
and
\begin{equation}
    d_{Z,X} = \sum_{c\in X} d_{Z,c}. \label{zx}
\end{equation}

Since we are using a consistency model, we have that
\begin{align}
    d_{Y,X} &= \sum_{c\in X} \left( d_{Z,c} - d_{Z,Y}\right) \notag \\
            &= \sum_{c\in X} d_{Z,c} - \sum_{c \in X} d_{Z,Y} \notag \\
            &= d_{Z,X} - |X| d_{Z,Y}, \label{consistency1}
\end{align}
where the last line is obtained by using Equation \eqref{zx}.

Applying consistency to the left-hand side of Equation \eqref{consistency1}, we have
\begin{align}
    d_{Z,X} - d_{Z,Y} = d_{Z,X} - |X| d_{Z,Y}.
\end{align}
This implies that $d_{Z,Y} = 0$ and thus treatments $Z$ and $Y$ are not distinct, which proves the uniqueness of the anchor treatment by contradiction.

\section{Impact of the Anchor in Additivity}\label{sec:impact}

The anchor treatment has been overlooked in the literature, but plays an important role in CNMA models. In this section, we show how different anchors used in the additivity equation \eqref{eq:additive} can impact the relative effects and treatment rankings for multicomponent treatments, emphasizing why it is important to use an anchor which agrees as much as possible with the data generating mechanism.

Consider a situation where there are 5 treatments available for a certain disease, $A, B, \mathcal{C}, \mathcal{D}$, and $E$, and for simplicity assume that a positive response is considered good. Also, assume that there is a network which contains treatments $A, B, \mathcal{C}, \mathcal{D}, E$, and $\mathcal{C+D}$. Assume that the effects of the individual treatments can be ordered as follows:
\[
E < \mathcal{D} < \mathcal{C} < B < A,
\]
that is, A has the largest effect, E is the smallest, and  $\mathcal{D}$, $\mathcal{C},$ and $B$ are somewhere in the middle.

Suppose that additivity is assumed with treatment $E$ as the anchor. This may correspond to using a placebo as the anchor, since it is less effective than other treatments. Then we have the following ordering of the relative effects with comparator $E$:
    \[
    d_{E,E} = 0 < d_{E, \mathcal{D}} < d_{E,\mathcal{C}} < d_{E,B} < d_{E, A}
    \]
    Computing the relative effect of $\mathcal{D+C}$ compared to E by applying additivity,
    \[
    d_{E,\mathcal{D+C}} = d_{E,\mathcal{D}} + d_{E, \mathcal{C}} > d_{E, \mathcal{C}}
    \]
    Therefore, under additivity with E as the anchor, the multicomponent treatment is implied to have a greater effect than either of its components alone. In other words, when each individual component of a multicomponent treatment is more effective than the anchor, the model expects the multicomponent treatment to be more effective than each of its components.
    
On the other hand, consider using B as the anchor for additivity. This could correspond to using a standard drug as the anchor. Then we have the following ordering of the relative effects with B as the comparator:
    \[
    d_{B,E} < d_{B,\mathcal{D}} < d_{B,\mathcal{C}} < d_{B, B} = 0 < d_{B,A}
    \]
    Computing the relative effect of $\mathcal{D+C}$ compared to B,
    \[
    d_{B,\mathcal{D+C}} = d_{B,\mathcal{D}}+ d_{B, \mathcal{C}} < d_{B,\mathcal{D}}
    \]
    so this additivity assumption implies that the multicomponent treatment $\mathcal{C+D}$ has less of an effect than either of its components alone. In other words, when each individual component of a multicomponent treatment is less effective than the anchor, the multicomponent treatment is expected to be less effective than each of its components.

Another issue arises when the modeller chooses an anchor which is a component in multicomponent treatments, which may seem like a reasonable choice to make. However, returning to our example, consider letting $\mathcal{D}$ be the anchor, and notice that $\mathcal{D}$ appears as a component in the multicomponent treatment $\mathcal{C+D}$.

We compute that the relative effects are 
\[
    d_{\mathcal{D}, E} < d_{\mathcal{D}, \mathcal{D}} = 0 < d_{\mathcal{D}, \mathcal{C}} < d_{\mathcal{D}, B} < d_{\mathcal{D}, A}
    \]
and by applying additivity we find that
\[
    d_{\mathcal{D}, \mathcal{D+C}} = d_{\mathcal{D}, \mathcal{D}} + d_{\mathcal{D}, \mathcal{C}} = 0 + d_{\mathcal{D}, \mathcal{C}} = d_{\mathcal{D}, \mathcal{C}},
    \]
since the relative effect of any treatment compared to itself is zero. Using the component $\mathcal{D}$ as the anchor forces $\mathcal{C+D}$ to have the same relative effect as $\mathcal{C}$ alone, as though adding the two treatments together does not have an effect. Although this implication may be reasonable in some scenarios, it seems a strong assumption to make without empirical evidence.

This example illustrates how three different anchors give three different models. The anchor which is best depends on the data. If the data shows that combining C and D has a beneficial effect, then the model using treatment E as the anchor will likely fit the data better. On the other hand, if combining treatments C and D produced a worse effect in the data, then the model using treatment B as the anchor will be more suitable. Finally, using treatment $D$ as the anchor may give the best fit if combining components does not have any effect. It is difficult to say what treatment is the most suitable as the anchor before fitting the models. Therefore, unanchored models are attractive because they do not require the anchor to be specified a priori.

\section{Proof of Equivalency of the Novel Models}\label{sec:proof}

Applying ${\Bi}^*$ to a vector of the observed log-odds ratios in each arm of study $i$ will give the data vector $\yyi^*$. More specifically, $U_{i,jk}^* = 1$ if the $k^\text{th}$ arm is the minuend of the $j^\text{th}$ contrast, $-1$ if the $k^\text{th}$ arm is the subtrahend in the $j^\text{th}$ contrast, and $0$ otherwise. The rows of ${\Bi}^*$ sum to zero and the first column of ${\Bi}^*$ (the common subtrahend) contains all $-1$'s. We can show that multiplying ${\Bi}^*$ by equation \eqref{eq:armmodel} gives equation \eqref{eq:contrastmu}:
\begin{align*}
    {\Bi}^*\logit(\pp_i) &= \alpha_i {\Bi}^* \one + {\Bi}^*\Ai \dd + {\Bi}^* \epsi\\
    &= {\Bi}^*\Ai \dd + \epsi^*\\
    &= \mmu_i^*,
\end{align*}
noting that ${\Bi}^*\one = \zero_{(a_i-1)\times 1}$, and ${\Bi}^*\epsi$ has the same probability distribution as $\epsi^*$. 
The latter follows from $\Bi{}^{*} \epsi \sim N(\zero, \Bi{}^{*} \Sigma_i{\Bi{}^{*}}')_{(a_i-1)\times 1}$, 
where $\Bi{}^{*} \Sigma_i{\Bi{}^{*}}' = 0.5\sigma^2 \Bi{}^{*} (J + I){\Bi{}^{*}}' = 0.5\sigma^2\Bi{}^{*} I{\Bi{}^{*}}' = \Sigma_i^{*}$, 
where $J$ is a $a_i\times a_i$ matrix of ones and $I$ is the $a_i\times a_i$ identity matrix.

\section{Data Generation Mechanism}\label{sec:dgm}

We adopt a data generation mechanism which produces arm-level data, and then transform the resulting arm-level data to contrast-based summaries. See Appendix \ref{sec:pars} for details on how the relative effect parameters are set. After setting the parameter values and the network structure, datasets are generated from the model in Subsection \ref{sec:welton}. First, $\logit(\pp_i)$, $i = 1, \dots, I$ is drawn from equation \eqref{eq:weltonmodel} and then each binomial probability $p_{ij}$ is determined via the inverse logit function. Then binomial counts for each arm $r_{ij}$ are drawn from equation \eqref{eq:bin}. 

Once outcomes are generated for all trial arms, the data are transformed to contrast-level summaries. The appropriate summary for the contrast between arm 2 and arm 1 in trial $i$ is the log-odds ratio 
\begin{equation}
    y_{i}^* = \log \left(\frac{r_{i2}/(n_{i2}-r_{i2})}{r_{i1}/(n_{i1}-r_{i1})}\right)
\end{equation}
with corresponding standard error
$SE_{i2} = \sqrt{\frac{1}{r_{i2}} + \frac{1}{n_{i2}-r_{i2}} + \frac{1}{r_{i1}} + \frac{1}{n_{i1}-r_{i1}}}$.\citep{borenstein2011introduction}

\section{Parameter Calculation in Simulations}\label{sec:pars}

Here we describe the process by which we set the relative effect parameters. 
In each network and sample size setting, we simulate two datasets corresponding to two different anchors. Let $k$ and $k'$ represent the first and second anchor treatments which are to be used in a given network respectively. First, the relative effects of the single component treatments relative to the first anchor treatment $k$, $d_{k,j}$ are set. Then the relative effects for single components with the second anchor as the comparator, $d_{k',j}$, are determined according to the consistency equations, that is, $d_{k',j} = d_{k,j} - d_{k,k'}$. The relative effect parameter vector corresponding to $k$ as the anchor is $\dd^{(k)}$, a $(t-1)$-length vector containing the relative effects for all components (except $k$) compared to treatment $k$, and similarly for $\dd^{(k')}$.

Although they are not required to fit the model, the multicomponent relative effects are needed to compute bias and MSE in the simulation. The multicomponent effects relative to $k$ when $k$ is the anchor is given by the sum of its component effects relative to $k$, and similarly for when $k'$ is the anchor. For comparison purposes, we transform the multicomponent relative effects relative to $k'$ using the consistency equations so we can compare the multicomponent effects relative to $k$ when $k$ is the anchor to the multicomponent effects relative to $k$ when $k'$ is the anchor. That is, we compute $d_{k,j}$ and $d_{k',j}$ for all multicomponent treatments according to $d_{k,j} = \sum_{i \in j} d_{k,i}$ and $d_{k',j}= \sum_{i \in j} d_{k',i}$. Then we calculate the multicomponent effects relative to $k$ \textit{when k' is the anchor} according to the consistency equations, $d_{k,j} (\text{where } k' \text{ is the anchor)} = d_{k',j} - d_{k',k}$. The multicomponent relative effects for each network and anchor are shown in Table \ref{tab:networkparams}.

The baseline effect parameters $\alpha_i$ depend on which treatment is being used as the anchor, since in our data generating model they represent the response in trial $i$ to the anchor treatment. For simplicity, we assume that the baseline response in every trial for a given anchor is the same, that is, $\alpha_i = \alpha$ for all $i = 1,\dots,I$ and for a given anchor treatment. To set the parameters, a reasonable value is chosen for anchor $k$, and then the corresponding $\alpha$ for anchor $k'$ is computed according to $\alpha + d_{k,k'}$.

\bibliography{addrefs}

\begin{thebibliography}{10}
\providecommand \doibase [0]{http://dx.doi.org/}%

\bibitem{petropoulou2021review}
Petropoulou M, Efthimiou O, R{\"u}cker G, et al. A review of methods for
  addressing components of interventions in meta-analysis. {\it PLoS One}
  2021\string; 16(2)\string: e0246631.

\bibitem{welton2009mixed}
Welton NJ, Caldwell D, Adamopoulos E, Vedhara K. Mixed treatment comparison
  meta-analysis of complex interventions: psychological interventions in
  coronary heart disease. {\it Am J Epidemiol} 2009\string; 169(9)\string:
  1158--1165.

\bibitem{rucker2019network}
R{\"u}cker G, Petropoulou M, Schwarzer G. Network meta-analysis of
  multicomponent interventions. {\it Biom J} 2020\string; 62(3)\string:
  808--821.

\bibitem{efthimiou2022bayesian}
Efthimiou O, Seo M, Karyotaki E, et al. Bayesian models for aggregate and
  individual patient data component network meta-analysis. {\it Statistics in
  medicine} 2022.

\bibitem{mills2011pharmacotherapies}
Mills EJ, Druyts E, Ghement I, Puhan MA. Pharmacotherapies for chronic
  obstructive pulmonary disease: a multiple treatment comparison meta-analysis.
  {\it Clin Epidemiol} 2011\string; 3(1)\string: 107.

\bibitem{mills2012calculating}
Mills EJ, Thorlund K, Ioannidis JP. Calculating additive treatment effects from
  multiple randomized trials provides useful estimates of combination
  therapies. {\it J Clin Epidemiol} 2012\string; 65(12)\string: 1282--1288.

\bibitem{madan2014synthesis}
Madan J, Chen YF, Aveyard P, et al. Synthesis of evidence on heterogeneous
  interventions with multiple outcomes recorded over multiple follow-up times
  reported inconsistently: a smoking cessation case-study. {\it J R Stat Soc
  Ser A Stat Soc} 2014\string; 177(1)\string: 295--314.

\bibitem{caldwell2016approaches}
Caldwell DM, Welton NJ. Approaches for synthesising complex mental health
  interventions in meta-analysis. {\it Evid Based Ment Health} 2016\string;
  19(1)\string: 16--21.

\bibitem{pompoli2018dismantling}
Pompoli A, Furukawa TA, Efthimiou O, Imai H, Tajika A, Salanti G. Dismantling
  cognitive-behaviour therapy for panic disorder: a systematic review and
  component network meta-analysis. {\it Psychol Med} 2018\string;
  48(12)\string: 1945--1953.

\bibitem{freeman2018component}
Freeman SC, Scott NW, Powell R, Johnston M, Sutton AJ, Cooper NJ. Component
  network meta-analysis identifies the most effective components of
  psychological preparation for adults undergoing surgery under general
  anesthesia. {\it J Clin Epidemiol} 2018\string; 98\string: 105--116.

\bibitem{dias2014generalised}
Dias S, Welton NJ, Sutton AJ, Ades A. {\it A generalised linear modelling
  framework for pairwise and network meta-analysis of randomised controlled
  trials}.
\newblock London, UK: National Institute for Health and Care Excellence (NICE)
  London .
\newblock 2014.

\bibitem{dias2018network}
Dias S, Ades AE, Welton NJ, Jansen JP, Sutton AJ. {\it Network meta-analysis
  for decision-making}.
\newblock Hoboken, NJ: Wiley .
\newblock 2018.

\bibitem{rucker2012network}
R{\"u}cker G. Network meta-analysis, electrical networks and graph theory. {\it
  Res Synth Methods} 2012\string; 3(4)\string: 312--324.

\bibitem{thorlund2012stability}
Thorlund K, Mills E. Stability of additive treatment effects in multiple
  treatment comparison meta-analysis: a simulation study. {\it Clin Epidemiol}
  2012\string; 4(1)\string: 75--85.

\bibitem{rucker2020component}
R{\"u}cker G, Schmitz S, Schwarzer G. Component network meta-analysis compared
  to a matching method in a disconnected network: A case study. {\it Biom J}
  2020\string; 63(2)\string: 447--461.

\bibitem{netmeta}
Rücker G, Krahn U, König J, Efthimiou O, Schwarzer G. {\it netmeta: Network
  Meta-Analysis using Frequentist Methods}. ; :  2020.
\newblock R package version 1.2-1.

\bibitem{rucker2014reduce}
R{\"u}cker G, Schwarzer G. Reduce dimension or reduce weights? Comparing two
  approaches to multi-arm studies in network meta-analysis. {\it Stat Med}
  2014\string; 33(25)\string: 4353--4369.

\bibitem{rucker2015ranking}
R{\"u}cker G, Schwarzer G. Ranking treatments in frequentist network
  meta-analysis works without resampling methods. {\it BMC Med Res Methodol}
  2015\string; 15(1)\string: 58.

\bibitem{salanti2011graphical}
Salanti G, Ades A, Ioannidis JP. Graphical methods and numerical summaries for
  presenting results from multiple-treatment meta-analysis: an overview and
  tutorial. {\it J Clin Epidemiol} 2011\string; 64(2)\string: 163--171.

\bibitem{spiegelhalter2002bayesian}
Spiegelhalter DJ, Best NG, Carlin BP, Van Der~Linde A. Bayesian measures of
  model complexity and fit. {\it J R Stat Soc Series B Stat Methodol}
  2002\string; 64(4)\string: 583--639.

\bibitem{denwood2016runjags}
Denwood MJ. runjags: An R package providing interface utilities, model
  templates, parallel computing methods and additional distributions for MCMC
  models in JAGS. {\it J Stat Softw} 2016\string; 71(1)\string: 1--25.

\bibitem{borenstein2011introduction}
Borenstein M, Hedges LV, Higgins JP, Rothstein HR. {\it Introduction to
  meta-analysis}.
\newblock West Sussex, UK: Wiley .
\newblock 2011.

\end{thebibliography}

\end{document}